# *The impact of field shape optimization: A feasibility study

Moorthy Muthuswamy and Yeh-Chi Lo

Department of Radiation Oncology, Mount Sinai School of Medicine, 1 Gustav L. Levy Place, New York, NY 10029, USA

## Abstract

The impact of field shape optimization is studied for prostate type geometry. For this study, 76 and 81 Gy plans were generated. Dose distributions for wedged plans and Intensity Modulated (IM) plans for three and seven fields were compared for a quadratic cost function. For wedged plans, a Simulated Annealing Algorithm (SAA) was used to optimize gantry angles, wedge angles, beam weights and field shapes. Two kinds of wedged plans were generated: 1) field sizes were determined by the requirement of full target coverage in the beam's-eye-view (fixed fields) and 2) the field shape, in particular at the critical organ target overlap region was also among the variables optimized. For IM plans the SAA was used to optimize gantry angles and a conjugate gradient algorithm was used to optimize the IM beam fluences. Both the field shape optimized wedged plans and IM plans had significantly superior dose area histograms of the target, rectum and the bladder and cost function values compared to the fixed field optimized wedged plans.

## Introduction

The primary task in radiation therapy treatment planning and delivery of radiation dose is to deliver maximum dose to the tumor while minimizing dose to the critical normal tissue surrounding it. Quantitative dose volume constraints on both tumor and the normal tissue(s) are normally part of a physician's prescription.

This task in reality depends on a number of variables that all affect the resulting radiation dose distributions in various ways. The following is a list of such variables:

1) number of radiation beams
2) orientation of the radiation beams
3) beam modifiers that are called "wedges" with which the beam intensity gradient in one of the dimensions of the beam cross section can be changed
4) the modality/energy of the radiation used (electron beam or photon beam and the energy of such beams)
5) beam weights
6) the shape of the cross sectional area of the beam
7) the beam fluence variation within the cross section of the beam, also called Intensity Modulation (IM)

In most of the institutions adjusting the variables listed from 1-6 is carried out by trial and error manually until dose distributions close to matching the prescription are found. This limits the ability to generate the best dose distribution and also necessarily makes the prescription requirements less complex.

This task of generating best dose distributions can also be viewed as an optimization problem. The prescription dose requirements on the tumor and normal tissue can be expressed in terms of cost functions (Webb 1997). The goal here is to optimize the cost function that depends upon a number of variables (above list 1-7). The optimization of variables 1-5 in the list may be called conventional optimization. Optimization of variables 1-6 may be called Field Shape Conventional Optimization (FSCO), the new technique proposed here. Optimization of variables 1-7, excluding #3, #5 and #6 in literature is known as Intensity Modulated (IM) Optimization. Optimization of wedge angles and beam weights has been studied before and authors point out the usefulness this optimization procedure (Oldham *et al* 1995). However, the fundamental problem in conventional optimization is its inability achieve desirable dose distributions in situations of significant tumor-dose limiting critical organ overlap (Webb 97). In this paper, such overlap situations either correspond to a critical organ lying inside the concave outline of the tumor or the tumor lying inside the concave outline of a critical organ. The problem of delivering dose with conventional approach in such situations is the following. While trying to ensure minimum prescription dose requirements on tumor at the same time leads to large dose being delivered to the part of the critical organ located in the overlap region. Some such situations occur in prostate tumor-rectum overlap, prostate tumor-bladder regions, tumors near spinal cord such as thyroid cancers, laryngeal tumors with subglottic extension, and hypopharyngeal cancers with postcricoid extension, and optic nerve sheath meningioma's.

Intensity Modulation Optimization

Intensity Modulation optimization promises to provide more improved dose distributions even in situations of significant tumor-dose limiting critical organ overlap (Webb 1997). To quote Webb (p 135, 1997), "... .prostate wraps around the rectum and is in turn itself wrapped around by part of the bladder, creating a requirement for IMB (Intensity Modulated Beam) therapy". Numerical back projection methods and simulated annealing based methods and others have been used for both dose and biology based cost functions in intensity modulation optimization (Brahme *et al* 1982, Brahme 1988, Webb 1989, Holmes *et al 1991,* Webb 1992, Kallman *et al 1992,* Wang *et al I996).*

In general, the number of degrees of freedom associated with this entire list 1-7, (excluding 3, 5 & 6) is large and even optimizing a limited number of the variables while including scatter dose can be complex and time consuming (Stein *et al* 1997). For instance, at Memorial Sloan Kettering Cancer the optimization problem is performed (Ling *et al 1996)* by fixing the number of beams and their orientations while allowing for beam fluence optimization within the cross section of the beam (IM, #7 on the above list). Even this is a multivariable complex optimization problem. This can be seen from the fact that if the cross section of the beam is divided into smaller grid sizes, in the 2-D space, the number of such grids is proportional to the square of the grid width. It is clear therefore that IM dose distributions couldn't be achieved manually by trial and error.

Simulated Annealing Algorithm is particularly suited to application in optimization problems that have cost functions with multiple minimums (Webb 1989, Mageras and Mohan 1992, Rosen 1994). In case of a cost function expressed in terms of Normal Tissue Complication Probabilities (NTCP) and Tumor Control Probabilities (TCP) multiple minimums are possible (Mageras and Mohan 1992). A cost function expressed as a function of squares of quadratic dose differences has only one minimum for fixed beam orientations. However when beam orientations are also optimized it may have multiple minimums (Bortfeld *et al I993).* In the recent years some groups have increased the complexity of the IM optimization problem by also optimizing beam orientations (Stein *et al* 1997) in simulations. Due to its complexity clinical applications of IM dose delivery involving beam orientation optimization is not popular. It has been observed that the beam orientations are important when the number of beams are small (Stein *et al* 1997).

Once the simulations are completed the IM optimized beam dose fluences at appropriate beam orientations need to be delivered to the prescribed site of a patient. There are several ways of delivering the IM optimized beam. They are: 1) NOMOS PEACOCK (Carol *et al 1995)* and Tomotherapy (Mackie *et al 1993)* systems based on the concept of binary temporal modulation; 2) overlapping fields with static delivery (Galvin *et al 1992* Bortfeld *et al* 1994, Boyer 1994, Gustafsson *et al 1995)* 3) sweeping variable gap with dynamic delivery (Kallman *et al 1988,* Convery *et al 1992,* Bortfeld *et al 1994,* Mohan 1996); 4) arc therapy with dynamic delivery (Yu 1995) and 5) construction of 2-D compensators (Djordjevich *et al I990).*

In general, these approaches of delivering dose to patients that use cutting edge technology are complex, expensive and time consuming in general compared to the conventional techniques (Ling *et al 1996,* Webb 1997). In many cases it requires buying expensive specialized technology such NOMOS PEACOCK (Carol *et al* 1995) and Tomotherapy (Mackie *et al 1993)* systems. Due to the complicated dose delivery scheme of IM there is also a problem of verification of its accuracy (Webb 1997). Ling *et al* (1996) have put forward an extensive QA system for dynamic leaf therapy. The verification and the QA issues related to the IM are considerably more complex than conventional therapy (that may include static MLC).

The proposed technique of Field Shape Conventional 0lJtimization (FSCO) plans:

Normally in the conventional optimization procedure of the cost function, the field shapes are chosen so that in the beam's-eye-view (i.e. from the direction of the beam) the field shape covers the entire planning target volume (target = tumor + margins representing clinical and setup uncertainties). We are proposing an approach of optimizing this field shape. This means in certain situations in the beam's-eye-view the target may not be covered. The Field Shape Conventional Optimized plans (FSCO) we propose are those in which the variables listed 1-6 in the previous page are optimized numerically. Since the last step listed before (#7) is not included, the IM is not performed in this optimization. Optimization of field shapes was investigated by Gustafsson *et al* (1995). However, their investigation does not simultaneously optimize wedge angles and gantry angles in addition to field edges. Only by optimizing the variables 1-6 on the list in previous page the potential of FSCO as an alternative to IM can be addressed.

Specifically, in this work we are optimizing the location of the field edge passing through the target-critical organ overlap region. As this problem is addressed in 2-D, field edge is optimized. In 3-D it would be optimizing field shapes. Field edge optimization is a subset of the more general field shape optimization. Field shape optimization of the regions of target not adjacent to the dose limiting normal tissues can also be done to provide adequate dose coverage to the target. Such an investigation was not carried out here.

Based upon the preliminary results (shown in the next section), we hypothesize that the FSCO plans are better than optimized conventional plans not involving field shape optimization (excluding #6 and 7 item on the list in the previous page). Based upon the preliminary results, the FSCO are also comparable to the generalized IM optimized plans (involving variables listed 1,2,4 and 7 in the previous page). These findings were observed for situations of tumor-dose limiting critical organ overlap. Since conventional optimization is a subset of FSCO optimization (this and the previous paragraphs) and based upon our preliminary findings, the FSCO optimization should very likely lead to better dose distributions than conventional optimization.

In this work conventionally optimized plans are compared with FSCO plans. This is done to study the impact of field shape optimization. These two are also compared with IM plans, as IM optimization remains the benchmark with which newer techniques can be compared. These comparisons were performed for three and seven field configurations for a prostate like geometry. A quadratic cost function used by Stein *et al* (1997) was optimized in this work.

## Method

In this study we seek to address the feasibility and impact of the FSCO technique. This can be realized in two dimensions. It is worth pointing out that the feasibility and the implications of intensity modulation were first shown in 2-D that used simple exponential dose decay type models (without scatter) by researchers before its extensive application in 3-D geometry (Webb 1992 and references therein). Extensive studies of FSCO plan's impact in three dimensions including scatter will be the subject of a future work. The comparison of these three approaches was performed for a 2-D phantom geometry of a half-ellipse of major and minor axes of 25.0 and 17.5 cm. The target (prostate + margins), rectum and the bladder were described by overlapping circles of radii 2, 2 and 2 cm for geometry I (figure la) and 2.6, 1.4 and 2 cm for geometry II (figure lb). The centers of the target and the rectum were separated by 3.5 cm and the centers of the target and the bladder were also separated by 3.5 cm. Two different relative sizes of the target and rectum were chosen to study dependence of the results on the degree of target-rectum overlap. Dose grid size of 2 mm was used. Tissue to Phantom Ratio (TPR) of a 15 MV beam was used to describe the attenuation of the beam. Inverse square law dose dependence was incorporated but not the beam divergence or the scatter. The dose distributions of individual beams used in this work have step function like fall of beyond the edge of field. This is due to not including scatter. In the case of FSCO plans only the field edge in the target-rectum overlap region were optimized. For instance, if the beam orientation corresponds to a gantry angle of 180 degrees (figure 1), then the field edges, as they do not lie in the target-rectum overlap region were not optimized. For a beam orientation corresponding to 90 degree gantry angle only the field edge lying in the target-rectum overlap region was optimized and not the field edge lying in the target bladder overlap region. As according to this cost function, bladder is not the primary dose-limiting organ due to the high threshold value for penalty for bladder dose (85 Gy) relative to tumor dose (81 Gy). So it was decided not to optimize to field-edge in the target-bladder overlap area. For the 76 Gy plan, bladder is not even a dose-limiting organ (Table 1). The need to spare sections of a dose-limiting organ from high dose becomes greater with a high target prescription dose. Hence we decided to make comparisons for two sets of dose plans (76 Gy and 81 Gy).

In the case of the conventional plan, gantry angles, wedge angles and beam weights were optimized using a Fast Simulated Annealing Algorithm (SAA) (Mageras and Mohan 1992). In the case of Field Shape Conventional Optimization (FSCO) plan, in addition to the gantry angles, wedge angles and beam weights, the location of the field edge at the rectum-target overlap region was also optimized using the SAA. Beam weight of a beam was taken to represent the dose from that beam at the isocenter. Wedge angles up to ±80 degrees were considered. Negative wedge angle here represents wedge oriented in the opposite direction. In the case of Intensity Modulated (IM) optimized plans, the gantry angles were chosen by the SAA while the IM beam dose fluence were optimized by a conjugate gradient algorithm (Press *et al* 1986). These are summarized in Table 1. For a given beam orientations (chosen by the simulated annealing algorithm) and for a quadratic cost function, IM fluence optimization has obviously one minimum. Hence to speed up the IM fluence optimization process the conjugate gradient algorithm was implemented. In general, more the number of beams is more conformal the dose distributions are. Since the cost function we employed does not penalize increasing number of beams, the number of beams was not optimized. We have generated dose distributions for 3 and 7 field plans. Data comparisons of two sets of field numbers should provide insights into the consistency of this optimization study.

Generally, not all of the gantry angles are treatable (Muthuswamy and Lam 1999). For ranges of gantry angles on either side of the couch part of the beam will pass through steel bars supporting the couch top. Such gantry angles are to be avoided. In this work, for the steel bar configuration of a Varian Clinac 2100C (Varian Corp., Palo Alto, CA, USA) machine and for this phantom geometry, the gantry angle ranges of (222.5-248.5) and (111.5-137.5) degrees were avoided as they lead to beam-couch intersection.

A quadratic cost function defined for a set of penalty factors (Stein *et al* 1997) was

$$F = \sum_{i=1}^{N} p_i (D_i - D_i^p)^2 / N$$

optimized. The cost function reflects certain prescription requirements on target and the normal tissues. This cost function has been found to be appropriate except for those treatment sites for which the tolerant dose of dose limiting normal tissues is low and the volume effect is large (Wang *et al 1995).* This cost function may be appropriate for this prostate dose study.

Here, N is the number of voxels, $D_i$ and $D_i^P$ are the relative values of actual and prescription (for prostate) or tolerance (for normal structure) doses at $i^{th}$ voxel. Also, $p_i$ is the penalty factor for $i^{th}$ voxel. The penalty factor for prostate is 1. The normal tissue penalty factors used for this cost function are given (Stein *et al* 1997) in Table 2. These penalty factors are different from zero only if the dose at a normal tissue voxel exceeds the tolerance dose given in the table. The reader is referred to Stein *et al* (1997) for details. During the IM fluence optimization process, the conjugate gradient (Press *et al* 1986) sometimes generated negative fluence entries. These were truncated to zero. This was similar to what was done by Stein *et al* (1997). Plans were normalized such that at about 95% of the target area receives the prescribed dose.

A Fast SAA described by Mageras and Mohan (1992) was used in this work. Like Stein *et al* (1997), we too found similar results with finite and zero temperatures in the simulated annealing optimization process. Local minimum trapping were avoided due to the long tail of the Cauchy distribution in the SAA (Mageras and Mohan 1992). The width of the Cauchy distribution was also updated in the manner described in Stein *et al* (1997) as,

$$W(t) = W_0/(1+t/R)$$

W(t) is the width of a variable that is varied in the SAA optimization process in the $t^{th}$ iteration. These variables were gantry and wedge angles, beam weights and field edge location. $W_0$ is the start width and R is a control parameter. A very large value of R decreases the step size. This however makes the convergence very slow. The value of the control parameter was adjusted through trail runs so that regardless of the value of the starting width $W_0$, approximately same value of the optimized cost function was reached. The Cauchy generator used in this algorithm was taken from a book by Rubinstein (1981). According to this algorithm, if x(t-1) is a variable such as a gantry angle in the (t-l) iteration step, for the width W(t), x(t) = x(t-l) + W(t)*A/B, where A and B are random variables belonging to the interval (-0.5, 0.5) and subject to the condition that $A^2 + B^2 \leq 0.25$. The VAX (Digital Corp., MA, USA) random generator function RAN was used as a random generator of variables A and B in the interval (0,1). Then a value of 0.5 was subtracted from A and B. The distribution of the x variable generator using this version of the Cauchy generator has the characteristic long tail indicative of a Cauchy distribution plotted in Mageras and Mohan (1992). Optimization was stopped when $(E-E_{min})/E < m$ and $W(t) < W_{stop}$, here E is the mean cost function for the last 5N (N = number of beams * number variables associated with each beam) and $E_{min}$ the minimum value of the cost function. The number of variables associated with SAA optimization for each beam: 1) in conventionally optimized plans were gantry and wedge angles and beam weight for a total of 3, 2) in FSCO plans were gantry and wedge angles, beam weight and field edge location for a total of 4, and 3) in IM plans only gantry angle was optimized through SAA for a total of 1. Typical values were m=0.01, $W_{stop}$(gantry angle)=1°, $W_{stop}$(wedge angle)=O.5°, $W_{stop}$(beam weight)=0.1 and $W_{stop}$(field edge location)=0.002 cm. With the grid size of 0.2 cm, $W_{stop}$ (field edge location) should be 0.2 cm and not 0.002 cm. This oversight led to increased time for the optimization process to converge. But this should not affect the results.

## Results

It must be kept in mind that what is considered, as a "better" plan is implicit in the way the cost function is defined here. The cost function (Stein *et al* 1997) used in this work is such that large area of rectum getting dose less than the tolerance dose was not penalized. But even a voxel of rectum area getting a dose above the tolerance (Table 2) has the cost function value increased by a penalty.

There is a left-right symmetry in both of the geometry. Due to that for each solution there is also another solution with the gantry angles mirror image of the displayed solution and appropriately oriented wedges. The optimized cost function values are displayed in the Table 3 for 3 and 7 field plans. In the SAA minimization process there is an uncertainty of about 10% in the converged value of the cost function from the global minimum value due to the finite number of iterations being performed. In this paper in addition to the comparisons between conventional optimization and FSCO, IM results are also shown with the possibility of establishing FSCO as an alternative to IM in some situations.

When field edge at the target-rectum overlap region was optimized, it was observed (Table 4) that the optimized half field edge width (denoted by ohwid) for some of the fields was smaller than the half field edge needed for full target coverage (denoted by fhwid). The values are displayed up to two decimal places in order to identify the voxel number they fall into. With a grid size of 2 mm, for geometry I, 81 Gy plan, field edge location of 1.80 cm is covered by 9 bins from the center and 1.85 cm is covered by 10 bins from the center. So even with a difference of only .05 cm, these two edge locations signify a difference. In the case of 76 Gy plan for geometry Ionly one gantry angle had a field edge located in the target-rectum overlap region and hence was optimized. The field edge at the target-bladder overlap region was not optimized.

Dose Area Histograms (DAH) of prostate target, rectum and the bladder are also shown in figures 2-7. The rectum DAHs are plotted in a log scale so that the high dose area are clearly distinguished for clarity. In figure 8 the beam fluence distribution of the IM beam profiles are shown for a 81 Gy plan. The dose fluence for all of bins displayed are defined as the dose deposited at the isocenter at a depth equaling the distance from the phantom surface to the isocenter for the central axis ray. The vertical dose profile distributions shown in figures 9 and 10 are useful in understanding advantages and limitations of various techniques compared here. The dose outside the target region in figures 9 and 10 arise from the intersection of beams (scatter is not included in this dose model).

In figures 11 and 12 gantry angle orientations of the three field arrangements are shown. In the case of conventional and FSCO plans the beam weights and wedge angles are also shown. The location of the field edges can be noted from Table 4. Due to so many beams involved the seven field configurations are not plotted. Relevant data are shown in Tables 5 and 6.

## Discussion

It is seen in Table 3 that the cost function values of FSCO plans are systematically smaller than conventionally optimized plans. The reason for the lower minimized value of the cost function is due to the optimization of the location of the field edge in the case of FSCO. From Table 4, it can be noted that in the case of FSCO, the field edge lies inside of the location one would expect based upon full target coverage in the beam's-eye-view. This analysis becomes clearer when dose profiles in figures 9 and 10 are studied. Clearly, the field edge has shifted in the case of FSCO compared to conventionally optimized plan. In addition in the case of FSCO the target dose is much. more uniform. In the IM fluence distributions shown in figure 8 the voxel located at the edge of the field (26) has lower value of intensity compared to voxels near the middle of the field. This voxel is at the target-rectum overlap region. This is a way in which high dose to the rectum in the IM approach is reduced in the optimization process.

The dose outside the target region for the seven fields plans in figures 9 and 10 arise from the overlap of exit or entrance part of the beams. For instance in figure 9a, the conventional plan has a larger bladder dose (in the region -3.6 to -2.6 cm) compared to the FSCO plan. From Tables 5 and Table 6, the gantry angles of the conventional plan with dominant beam weights are 140.6 and 205.7 degrees. These are orientations of gantry angles originating from the direction of the bladder (figure 1). Where as in the case of FSCO the gantry angles with dominant beam weights were 84.8 and 274.8 degrees. These are nearly lateral orientations, avoiding considerable portions of rectum and bladder.

The optimized beam orientations (gantry angles) are different between conventional FSCO plans (figures 11 and 12 and Table 5). At least in the case of three fields FSCO plans, there appear to be more inclination for the beams to come from the direction of the dominant dose limiting critical organ, rectum compared to FSCO. This is even more so with IM plans. This is consistent with the observations in the previous work of Stein *et al (1997)*.

From the DAHs in figures 2-5 it can be seen that in general, compared to conventionally optimized plans, both FSCO and IM optimized plans have more uniform target coverage and the maximum rectum dose is also reduced in these plans. While the cost function values for the 3 field FSCO and IM plans are not significantly different, for the 7 field plans they are different. But this is not reflected in the DAHs (figures 3a and 3b) significantly. So it appears that from the point of view of DAHs the FSCO appears to have done as well as the IM plans. Bladder DAHs for the FSCO and the IM 76 Gy plans are better than the conventionally optimized plans even though the bladder tolerant doses and penalty factors are not used for the 76 Gy plan. For the 81 Gy plan where these considerations are taken into account, the FSCO and IM plans still come out ahead compared to conventional plans.

The maximum target dose in figure 3b for the conventional plan is around 113 Gy, a rather high target dose. In this 81 Gy plan, there is a heavy penalty for getting rectum overdosed compared to target (Table 2). Therefore, a way to keep the cost low is through a higher target dose inhomogeneity. Another reason for the large dose uniformity is the requirement that about 95% of the target area gets the prescription dose. We believe this is what the optimization algorithm has done considering the consistency of both (high) 3 field and 7 field conventional cost function values (Table 3) compared to FSCO and IM plans. From the point of view of conventional planning this target non-uniformity associated with all of these plans may be regarded as a disadvantage with this type of cost function. However, the FSCO and IM plans have much better target uniformity with the same cost function.

In this work scatter and divergence were not taken into account, as was 3-D geometry. Since one is comparing different technique rather than absolute dose evaluations, the relative impact of including beam divergence is likely to be comparable. Hence the relative results between conventional optimization and FSCO shouldn't depend upon the divergence.

Scatter was not implemented in the dose calculation algorithm. The impact of scatter is transfer of the dose around the field edge and its deposition outside of the field margins (Mohan *et al* 1996). From figures 9 and 10, it can be seen that the relative impact of including scatter should be comparable. This should be because we have seen that reduced cost function value in the case of FSCO has come about through the reduced field width in the overlap region. Therefore, we conclude that the significant improvement we have seen with FSCO over conventionally optimized plan should remain when scatter is taken into account.

However, the inclusion of scatter should make IM plans better compared to the FSCO plans due to the ability of IM plans to modulate intensity around the field edges (Mohan *et a1 1996).* In particular this aspect of IM should lead to more uniform dose coverage of the target (Mohan *et al* 1996). In order to provide adequate dose coverage to the target, in 3-D it is conceivable that more beams may be required in the case of FSCO plans compared to the IM plans.

Arbitrary wedge angles are generated as part of this optimization process for non-IM plans. Arbitrary wedge angles could be achieved using a dynamic wedge. When 3-D application of this model are considered, constraints on more easily attainable wedge angles such as less than 60 degrees can be considered. The advantage of considering wedge angles less than a physically available wedge angle is that, any wedge angle less than a physical wedge angle could be achieved through a combination of that physical wedge and an open field.

Application of FSCO technique to the 3-D geometry requires addressing several issues. Among them being able to deliver arbitrary wedge angle beams in two directions. This could be achieved by using dynamic wedging in two directions or through collimator rotations of physical wedges. Also, the impact of defining average "wedge angle" in going from 2-D to 3-D needs to be investigated. Even in 2-D, optimizing field shape in more than one location needs to be investigated. The method of optimization of field shapes could be extended to 3-D (Gustafsson *et al 1995)* geometry similarly. Field shape optimization in 3-D geometry should translate to optimizing 2-D field shapes.

Since conventional optimization is a subset of FSCO, FSCO plans are always going to be better than conventionally optimized plans. The extent of improvement of FSCO over conventional plan needs to be investigated further for other types of geometry and cost functions. In the current algorithm variables are searched in a continuous space. We have seen that the impact of small changes in many of the parameterized variables on dose distributions is minimal, hence discrete sampling of variables can be employed. This should speed up the optimization process.

Since one would like to compare the "best" conventional plan with the "best" FSCO plan, the added complexity of optimizing gantry angles was taken. One possible study is to compare for a given gantry, wedge angles, beam weights and for a set of beam orientations, the impact of field shape optimization only. This is an ongoing study and will be part of another paper. As we have seen the "best" conventional, FSCO and IM plans may have different gantry angle orientations (figues 11 and 12 and Table 5). This may also imply that optimizing beam orientation in conjunction with field shape works to make the plans better. This is the reason for also optimizing gantry angles in this work. It is conceivable that in future gantry angles may also be routine optimized (Stein *et al* 1997). Another issue to address is whether simultaneous optimization of gantry angles and IM beam fluence can lead to different results. We believe that due to the quadratic form of the cost function used in this study, choosing the gantry angle first followed by IM optimization by the conjugate gradient algorithm leads to the global minimum. The consistency of the minimized cost function values of IM plans in Table 3 may tend to support this view. Also, without first defining the beam orientation by the selection of the gantry angle, solving IM fluence by conjugate gradient algorithm is not defined.

In a previous study (Mohan *et al 1996)* in which after the IM plans were optimized, the margins were adjusted when scatter was taken into account to arrive at "better" dose distributions. However, these authors do not seek to optimize the location of the margins (or equivalently the field shapes as they are called in this paper) with respect to the same cost function.

This study shows that it possible to improve conventionally optimized plans significantly by including field shape optimization. As we have seen, clearly, the feasibility and the impact of field shape optimization can be realized in 2-D. In FSCO plans, the field shape is the result of the optimization algorithm as opposed to full target coverage in the beam's-eye view of the conventional approach. So the delivery of field shape optimized plans should not be significantly different from conventional plan delivery. Even a mutileaf collimator may not be needed for the delivery of FSCO plans, as the fields may be shaped using cerrobend blocks. It probably should not take any more time to deliver FSCO treatments as compared to conventional treatments. It is important to note that conventional hardware can be used for treatment delivery of the FSCO, which can be carried out in most community hospitals.

The consistently close comparison between FSCO and IM techniques is worth noting. Importantly, this work also points to the potential of FSCO as an alternative to IM technique in some situations. Such situations are those of critical organ lying in the concave outlines of the tumor or vice versa (overlapping situations). In non-overlapping situations, Sherouse (1994) had pointed out that by optimizing the wedge angles one could achieve desired target dose distributions, especially for a small target. However, for a large target it may not be possible to achieve a desired dose distribution by adjusting beam weights, field shape, gantry and wedge angles globally (Webb 1997). In such situations IM may be a better options unless one opts for multi segmental treatment with FSCO. The FSCO technique needs to be further investigated with various shapes of critical organ-target overlap in 3-D with full-fledged dose models. These investigations need to be carried out with biologically based cost functions as well (Webb 1997).

**References**

Bortfeld T, Burkelbach J, Boesceke R and Schlegel W 1992 Three dimensional solution of the inverse problem in conformation radio-therapy in *Advanced radiation Therapy: Tumor response monitoring and treatment planning,* ed A Breit (Berlin: Springer) 649-53

Bortfeld T and Schlegel W 1993 Optimization of beam orientations in radiation therapy: some theoretical considerations *Phys. Med. Biol.* **38** 291-304

Bortfeld T, Boyer A L and Schlegel W *et al* 1994 Realization and verification of three dimensional conformal radiotherapy with modulated fields *Int. J. Radiat. Oncol. Biol. Phys.* **30** 899-908

Bortfeld T, Kahler D L, Waldron T J and Boyer A L *et all994* X-ray field compensation with multileaf collimators *Int. J. Radiat. Oncol. Biol. Phys.* **28** 723-30

Boyer A L 1994 Use of MLC for intensity modulation *Med. Phys.* **21** *1007*

Brahme A, Roos J and Lax I 1982 Solution of an integral equation encountered in radiation therapy *Phys. Med. Biol.* **27** 1221-29

Brahme A 1988 Optimization of stationary and moving beam radiation therapy techniques *Radiother. Oncol.* **12** 129-40

Carol M P 1995 PEACOCK: a system for planning and rotational delivery of intensity modulated fields *Int. J. Imag. Sys. Tech.* **6** 56-61

Convery D J and Rosenbloom M E 1992 The generation of intensity-modulated fields for conformal radiotherapy by dynamic collimation *Phys. Med. Biol.* **37** 1359-74

Djordjevich A, Bonham J, Hussein E M A, Andrew J Wand Hale M E 1990 Optimal design of radiation compensators *Med. Phys.* **17** 397-404

Galvin J M, Chen X G and Smith R M 1992 Combining multi leaf fields to modulate fluence distributions *Int. J. Radiat. Oncol. Biol. Phys.* **27** 697-705

Gustafsson A, Lind B K, Svensson Rand Brahme A 1995 Simultaneous optimization of dynamic leaf collimation and scanning patterns of compensating filters using generalized pencil beam algorithm *Med. Phys.* **22** 1141-56

Holmes T, and Mackie T R 1991 A unified approach to the optimization of brachytherapy and external beam therapy *Int. J. Radiat. Oncol. Biol. Phys.* **20** 859-73

Kallman P, Lind B K, Eklof A and Brahme A 1988 Shaping of arbitrary dose distributions by dynamic multileaf collimation *Phys. Med. Biol.* **33** 1291-300

Kallman P, Lind B K and Brahme A 1992 An algorithm for maximizing the probability of complication free tumor control in radiation therapy *Phys. Med. Biol.* **37** 871-90

Ling C C, Burman C and Chui C *et al* 1996 Conformal radiation treatment of prostate cancer using inversely-planned intensity modulated photon beams produced with dynamic multileaf collimation *Int. J. Radiat. Biol. Phys.* **35** 721-30

Mackie T R, Holmes T and Swerdloff S, *et al* 1993 Tomotherapy: a new concept for delivery of dynamic conformal therapy *Med. Phys.* **20** 1709-19

Mageras G and Mohan R 1992 Application of fast simulated annealing to optimization of conformal radiation treatments *Med. Phys.* **20** 639-47

Mohan R 1996 Intensity modulation in radiotherapy. In: J Palta, T R Mackie, eds *Teletherapy: past and future* (Madison: Advanced Medical Publishing) 761-92

Mohan R, Wu Q and Stein J 1996 Intensity modulation optimization, lateral transport of radiation and margins *Med. Phys.* **23** 2011-21

Muthuswamy M S and Lam K 1999 A method of beam-couch intersection detection *Med. Phys.* **26** 229-35

Muthuswamy M, and Forster K 1998 A comparison of field shape optimized conventional and intensity modulated prostate plans *Med. Phys.* **25** A96 (abstract at www.aapm.org)

Oldham M, Neal A J and Webb S 1995 The optimization of wedge filters in radiotherapy of the prostate *Radiother. Oncol.* **37** 209-20

Press W H, Teukolsky S A and Vetterling W T 1986 Numerical Recipes in Fortran: The Art of Scientific Computing *Cambridge University Press*

Rosen I 11994 Use of simulated annealing in conformal therapy optimization *Med. Phys.* ***21*** 1005

Rubinstein R 1981 Simulation and Monte Carlo Method *John Wiley and Sons 91-2*

Sherouse G W 1994 A simple method for achieving uniform dose from arbitrary arrangements of non-coplanar fixed fields *Int. J. Radiat. Biol. Phys.* **30** 240

Stein J, Mohan R, Wang X -H, Bortfeld T, Wu Q, Preiser K, Ling C F and Schlegel W 1997 Number and orientation of beams in intensity-modulated radiation treatments *Med. Phys.* ***24*** 149-60

Yu C X 1995 Intensity-modulated arc therapy with dynamic multileaf collimation: an alternative to tomotherapy *Phys. Med. Biol.* **40** 1435-49

Wang X -H, Mohan R, Jackson A, Leibel SA, Fuks Z and Ling C F 1996 Optimization of intensity modulated 3D conformal treatment based on biological indices *Radiother Oncol.* **37** 140-52

Webb S 1989 Optimization of conformal therapy treatments by simulated annealing *Phys. Med. BioI.* **34** 1349-69

Webb S 1992 Optimization by simulated annealing of three-dimensional conformal treatment planning for radiation fields defined by multi-leaf collimator: IT, Inclusion of two dimensional modulation of x-ray intensity *Phys. Med. Biol.* **37** 1689-704

Webb S 1997 *Physics of conformal therapy* (Bristol: Institute of Physics Publishing)

## Figure Legends

**Figures la and Ib:**

Here, a 2-D phantom geometry of a half-ellipse of major and minor axes of 25.0 and 17.5 cm is shown. The target (T), rectum (R) and bladder (B) are described by overlapping circles of radii 2, 2 and 2 cm for geometry I (figure 1a) and 2.6, 1.4 and 2 cm for geometry II (figure 1b) respectively. The centers of target and rectum and the target and bladder are separated by 3.5 cm.

**Figures 2a and 2b:**

Three field 81 Gy plan for geometry I and 76 Gy plan for geometry II are shown. DAHs of conventionally optimized, Field Shape Conventional Optimization (FSCO) and Intensity Modulated (IM) optimized plans are shown for comparison for target and rectum.

**Figures 3a and 3b:**

Seven field 76 Gy plan for geometry I and 81 Gy plan for geometry II are shown for prostate target and rectum.

**Figures 4-7:**

Seven field and three field bladder comparisons for 76 Gy and 81 Gy plans for geometry I & II.

**Figure 8:**

Intensity Modulation (IM) beam fluences for 3 field 81 Gy plan.

**Figures 9-10:**

Dose profiles through the axis (vertical, Y-axis) passing through the centers of rectum, target and bladder.

**Figures 11-12:**

For 3 fields plan, gantry and the wedge angle orientations and beam weights (in Gy) of the conventional and FSCO plans are shown. The orientations of IM plans are also shown for comparison.

**Table 1.** Variables optimized in different techniques.

| Technique | Variables optimized |
|---|---|
| Conventional optimization | beam weight<br>gantry angle<br>wedge angle |
| Field Shape Conventional Optimization (FSCO) | beam weight<br>gantry angle<br>wedge angle<br>field edge location |
| Intensity Modulation (IM) optimization | IM beam fluence<br>gantry angle |

**Table 2.** Tolerance doses($D^P$) and penalty factors(p) for normal tissues

| Organ | 76 Gy plan | 81 Gy plan |
|---|---|---|
| | *$D^P/p$* | *$D^P/p$* |
| Healthy Tissue[a] | | *90/2* |
| Rectum | *68/3* | *55/3* |
| Bladder | | *85/3* |

[a]A 1 cm ring of healthy tissue around the target is considered when optimum orientations of beams are determined, i.e. for the 81 Gy plans only. This is used as a means to avoid clustering of beams and thus hot spots outside the target.

**Table 3.** Minimized cost function values

| | Geometry I | | | | Geometry II | | | |
|---|---|---|---|---|---|---|---|---|
| | 76 Gy plan | | 81 Gy plan | | 76 Gy plan | | 81 Gy plan | |
| | 3 fld | 7 fld | 3 fld | 7 fld | 3 fld | 7 fld | 3 fld | 7 fld |
| Conven. Plan | 4.3 | 4.1 | 43.3 | 42.4 | 28.4 | 28.2 | 321.4 | 268.2 |
| FSCO plan | 0.75 | 0.54 | 4.6 | 4.1 | 0.67 | 0.61 | 5.4 | 5.3 |
| IM plan | 0.39 | 0.02 | 4.4 | 2.0 | 0.71 | 0.02 | 2.0 | 3.3 |

**Table 4.** A comparison of field edge values in cm for 3 field (FSCO) plans

| | Geometry I | | | | Geometry II | | | |
|---|---|---|---|---|---|---|---|---|
| | 76 Gy plan | | 81 Gy plan | | 76 Gy plan | | 81 Gy plan | |
| | ohwid | fhwid | ohwid | fhwid | ohwid | fhwid | ohwid | fhwid |
| Gantry angle I | 1.55 | 1.78 | 1.55 | 1.78 | 2.17 | 2.48 | 2.35 | 2.46 |
| Gantry angle II | | | 1.80 | 1.85 | 2.37 | 2.47 | 2.17 | 2.48 |

**Table 5.** Optimized gantry angles

| | Geometry I | | | | Geometry II | | | |
|---|---|---|---|---|---|---|---|---|
| | 76 Gy plan | | 81 Gy plan | | 76 Gy plan | | 81 Gy plan | |
| | 3 fld | 7 fld | 3 fld | 7 fld | 3 fld | 7 fld | 3 fld | 7 fld |
| Conven. | 89.4 | 69.9 | 87.3 | 20.8 | 84.9 | 50.0 | 141.3 | 55.4 |
| | 204.5 | 75.1 | 204.9 | 89.4 | 203.4 | 85.1 | 218.3 | 85.6 |
| | 277.7 | 199.8 | 274.3 | 144.4 | 210.9 | 140.6 | 269.6 | 150.8 |
| | | 207.2 | | 203.6 | | 141.4 | | 193.6 |
| | | 270.6 | | 208.4 | | 200.0 | | 203.3 |
| | | 284.9 | | 270.5 | | 205.7 | | 274.4 |
| | | 339.6 | | 291.8 | | 274.9 | | 304.6 |
| FSCO | 12.5 | 23.5 | 88.4 | 21.5 | 87.9 | 22.0 | 92.7 | 32.9 |
| | 88.1 | 83.1 | 200.9 | 52.0 | 92.8 | 84.8 | 207.8 | 88.2 |
| | 222.5 | 199.3 | 276.9 | 83.1 | 317.5 | 197.2 | 271.7 | 194.9 |
| | | 222.1 | | 89.4 | | 222.2 | | 222.3 |
| | | 267.4 | | 276.8 | | 270.3 | | 271.8 |
| | | 278.5 | | 352.4 | | 274.8 | | 275.0 |
| | | 351.6 | | 358.0 | | 280.6 | | 346.5 |
| IM | 89.6 | 42.6 | 195.5 | 41.9 | 65.8 | 89.8 | 83.1 | 55.7 . |
| | 277.0 | 89.2 | 272.6 | 72.1 | 90.4 | 92.7 | 275.7 | 78.6 |
| | 293.6 | 93.9 | 278.5 | 96.4 | 284.6 | 181.1 | 307.8 | 103.9 |
| | | 102.4 | | 207.3 | | 210.7 | | 182.0 |
| | | 169.7 | | 264.0 | | 257.1 | | 215.9 |
| | | 220.5 | | 298.9 | | 275.0 | | 274.3 |
| | | 341.9 | | 314.5 | | 277.9 | | 308.4 |

**Table 6.** Optimized beam weights and wedge angles for 7 field plans

| | Geometry I | | | | Geometry II | | | |
|---|---|---|---|---|---|---|---|---|
| | 76 Gy plan | | 81 Gy plan | | 76 Gy plan | | 81 Gy plan | |
| | beam weight | wedge angle | beam weight | wedge angle | beam weight | wedge angle | beam weight | wedge angle |
| Conven. | 3.95 | 14.7 | 11.1 | 80.0 | 0.02 | -6.5 | 16.8 | 78.7 |
| | 26.9 | 45.1 | 6.6 | 37.4 | 1.2 | 51.0 | 21.6 | 69.3 |
| | 4.59 | 0.55 | 1.6 | 26.3 | 34.5 | 43.0 | 8.8 | 78.6 |
| | 9.11 | 32.5 | 9.7 | -2.7 | 6.6 | 50.5 | 0.01 | 24.0 |
| | 7.7 | -9.4 | 9.58 | 8.6 | 6.5 | 80.0 | 9.3 | -42.9 |
| | 22.3 | -42.0 | 22.9 | -50.5 | 26.9 | 36.7 | 19.8 | -70.5 |
| | 2.25 | -40.8 | 21.9 | -62.0 | 6.1 | -40.0 | 16.2 | -78.5 |
| FSCO | 14.7 | -33.6 | 0.31 | 22.1 | 5.8 | 16.6 | 0.52 | 59.5 |
| | 15.0 | 23.4 | 16.2 | 66.9 | 23.0 | 39.2 | 24.0 | -27.6 |
| | 13.5 | -9.2 | 21.5 | -47.1 | 9.98 | -26.8 | 7.7 | -62.5 |
| | 9.7 | 15.9 | 30.7 | -51.0 | 5.1 | 6.1 | 17.3 | 40.3 |
| | 8.7 | 7.4 | 2.03 | 48.7 | 7.7 | 62.2 | 4.61 | -32.0 |
| | 9.9 | -27.3 | 3.75 | 17.8 | 19.3 | -20.4 | 22.2 | -5.9 |
| | 6.0 | 6.0 | 7.7 | -39.3 | 6.3 | 20.5 | 5.59 | 4.89 |

Fig. 1a: Geometry I

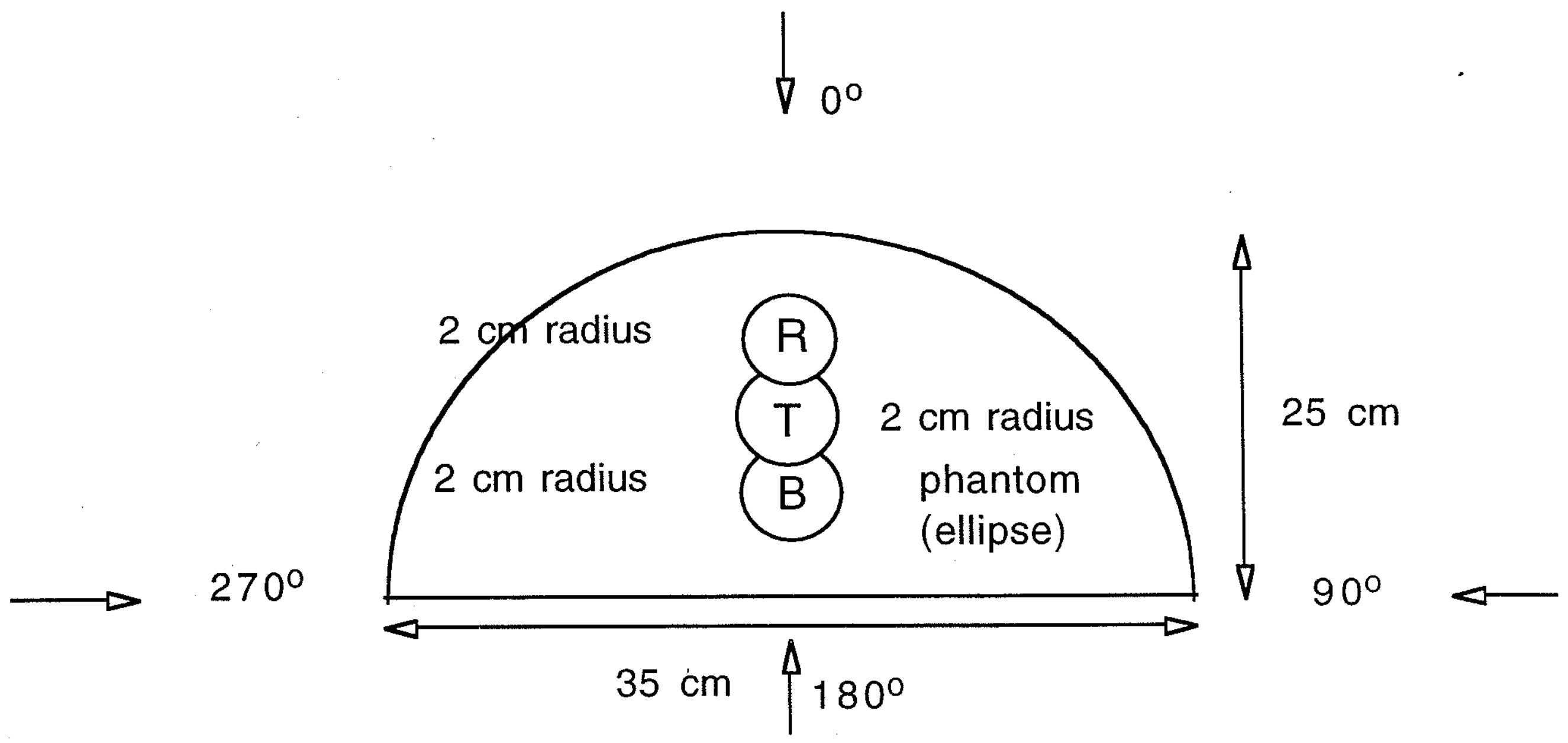

Fig. 1b: Geometry II

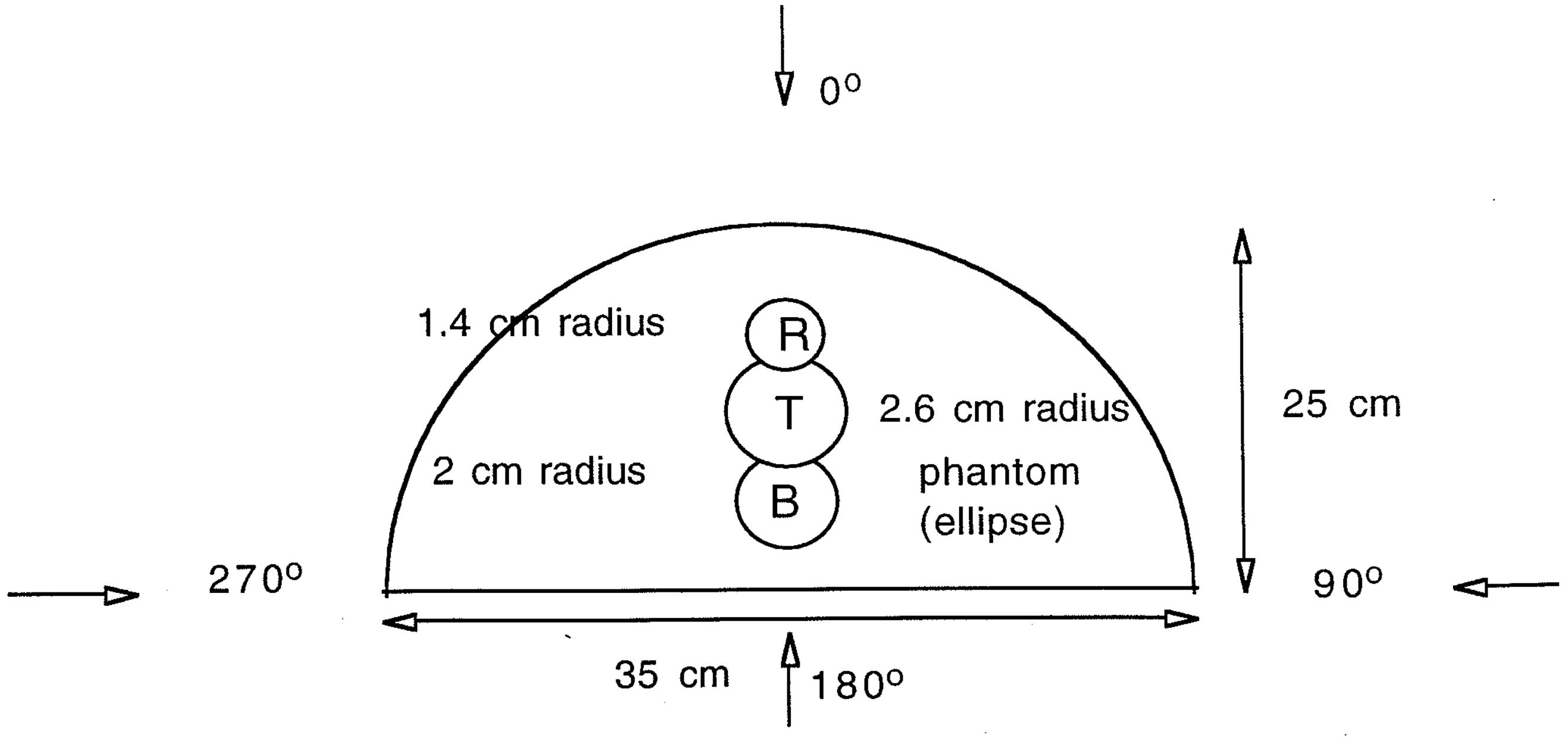

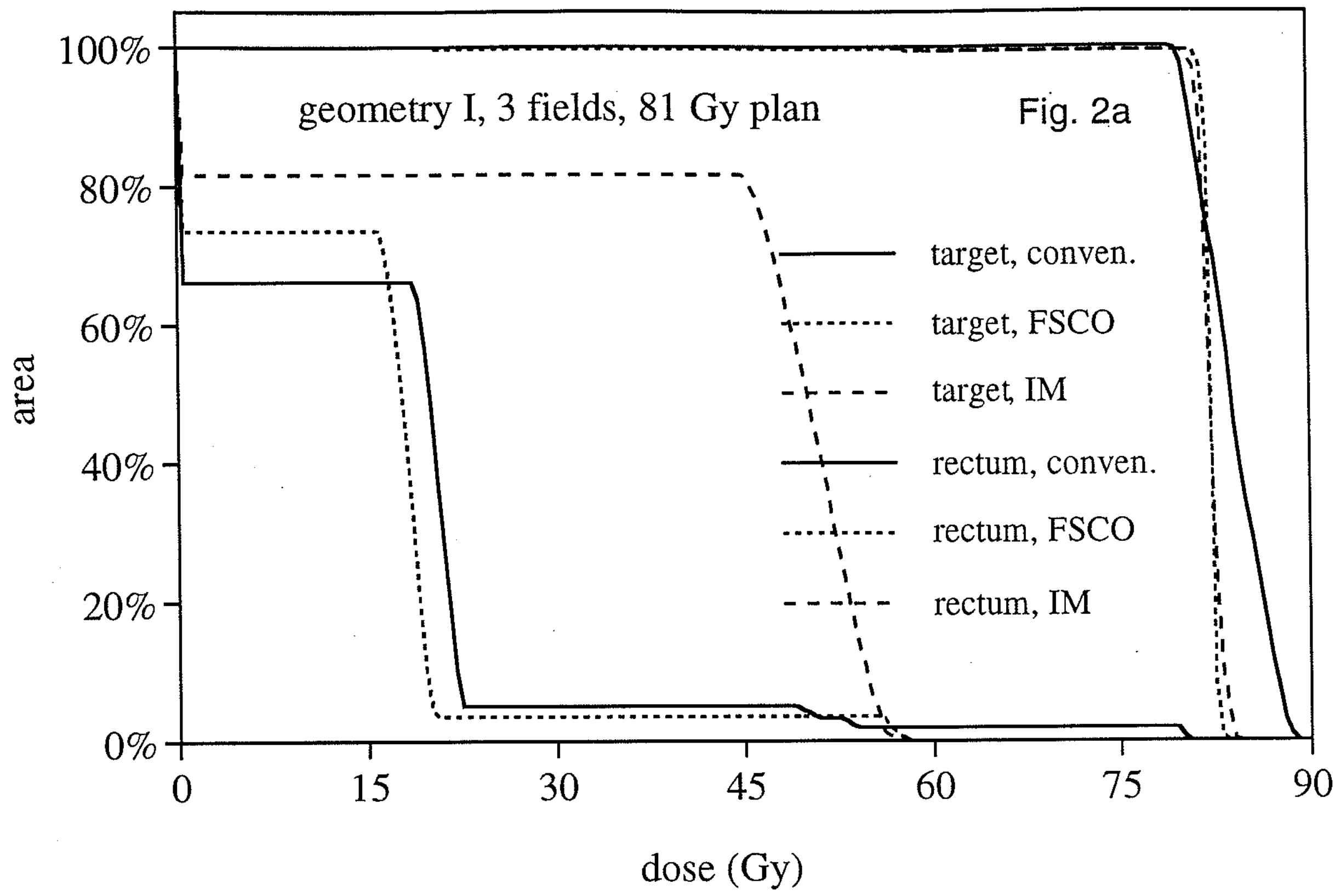

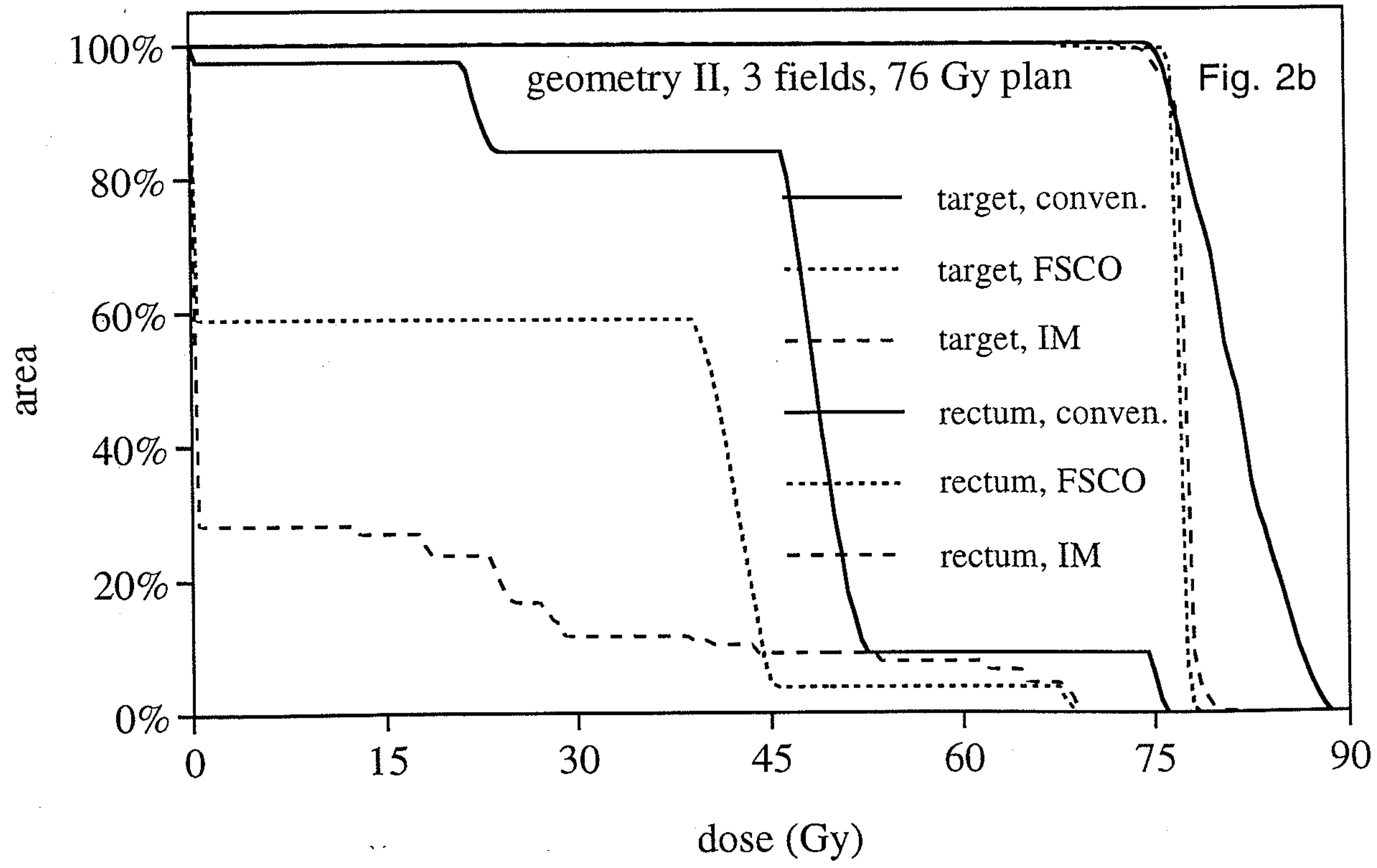

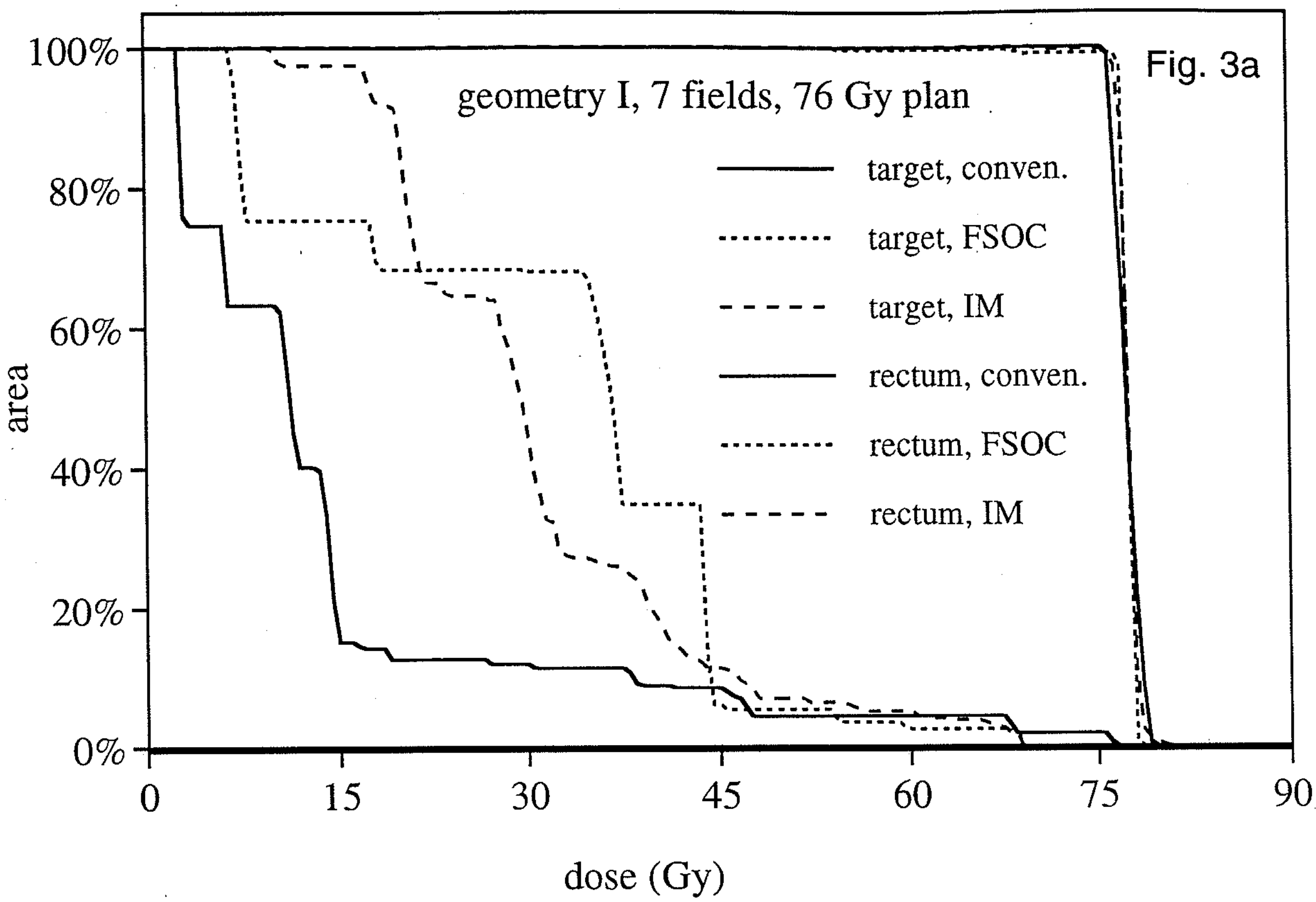

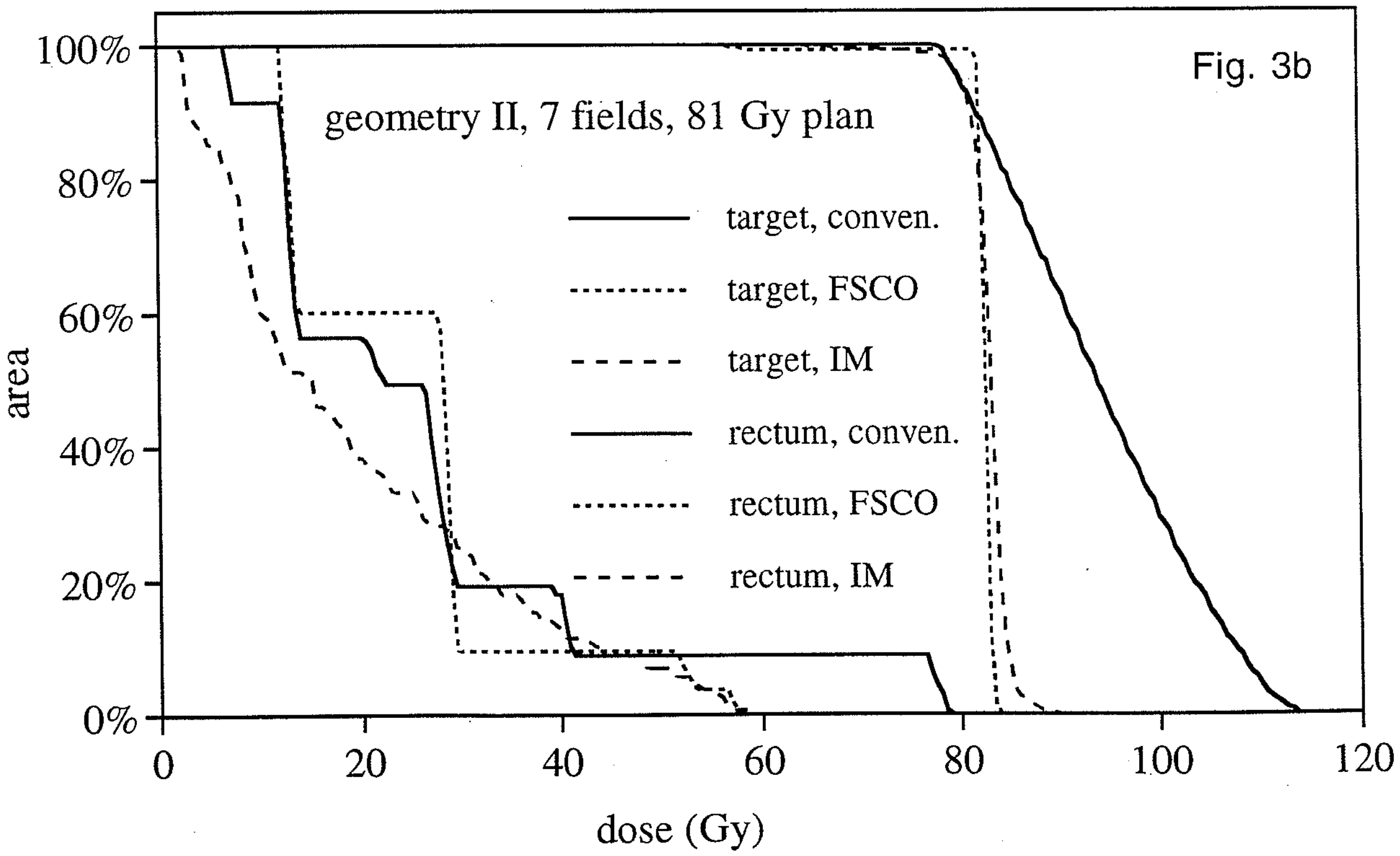

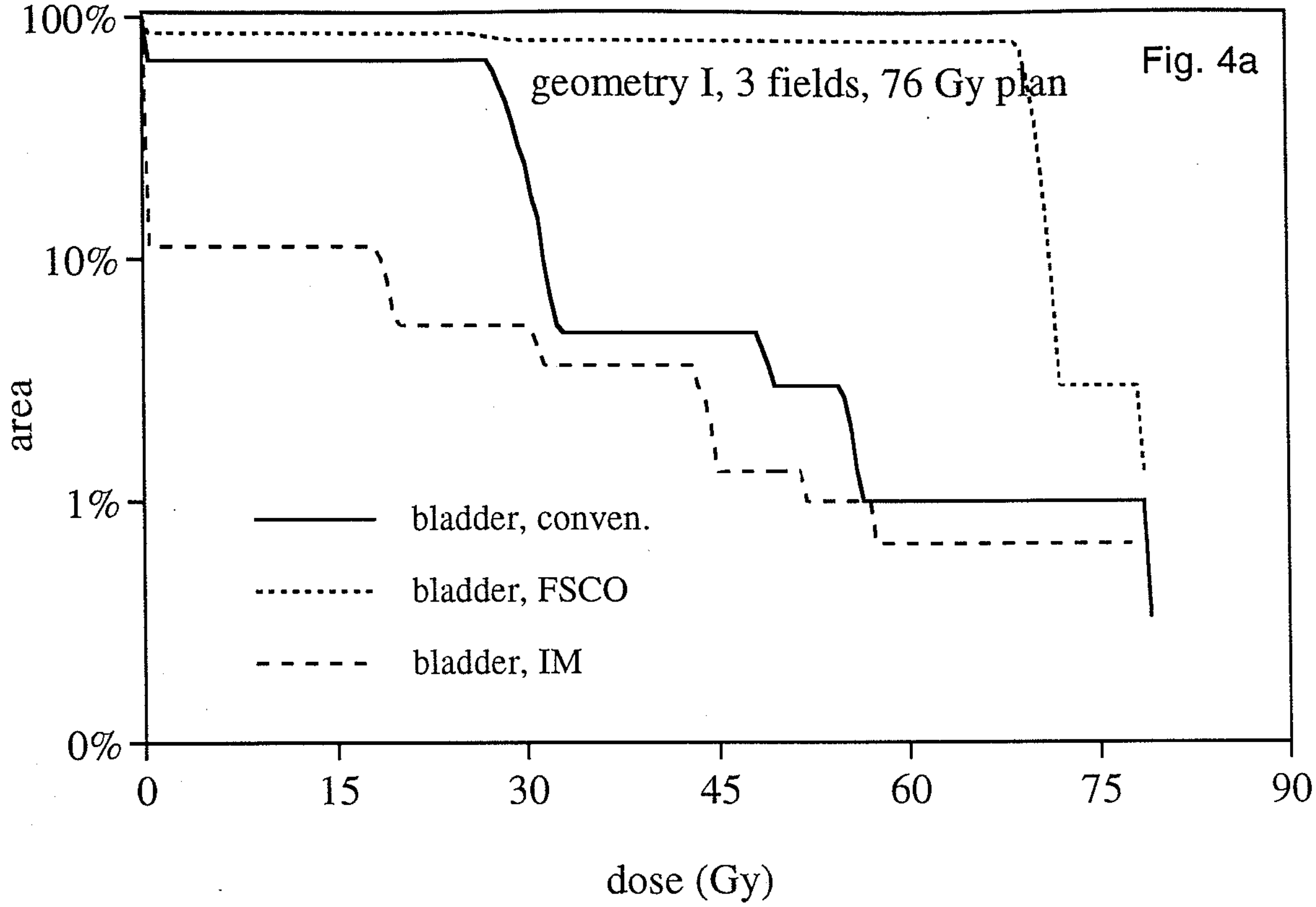

Fig. 4a

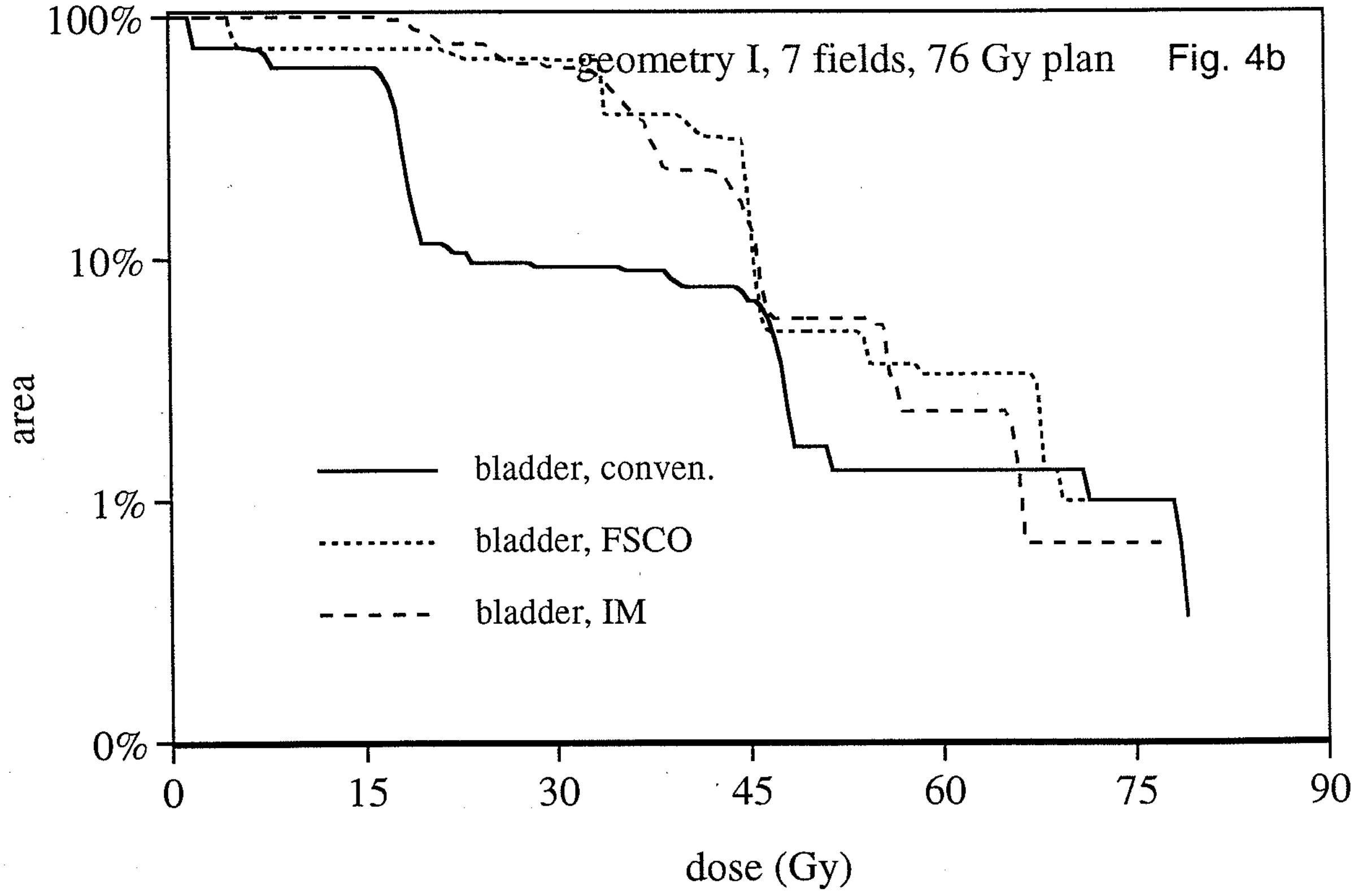

Fig. 4b

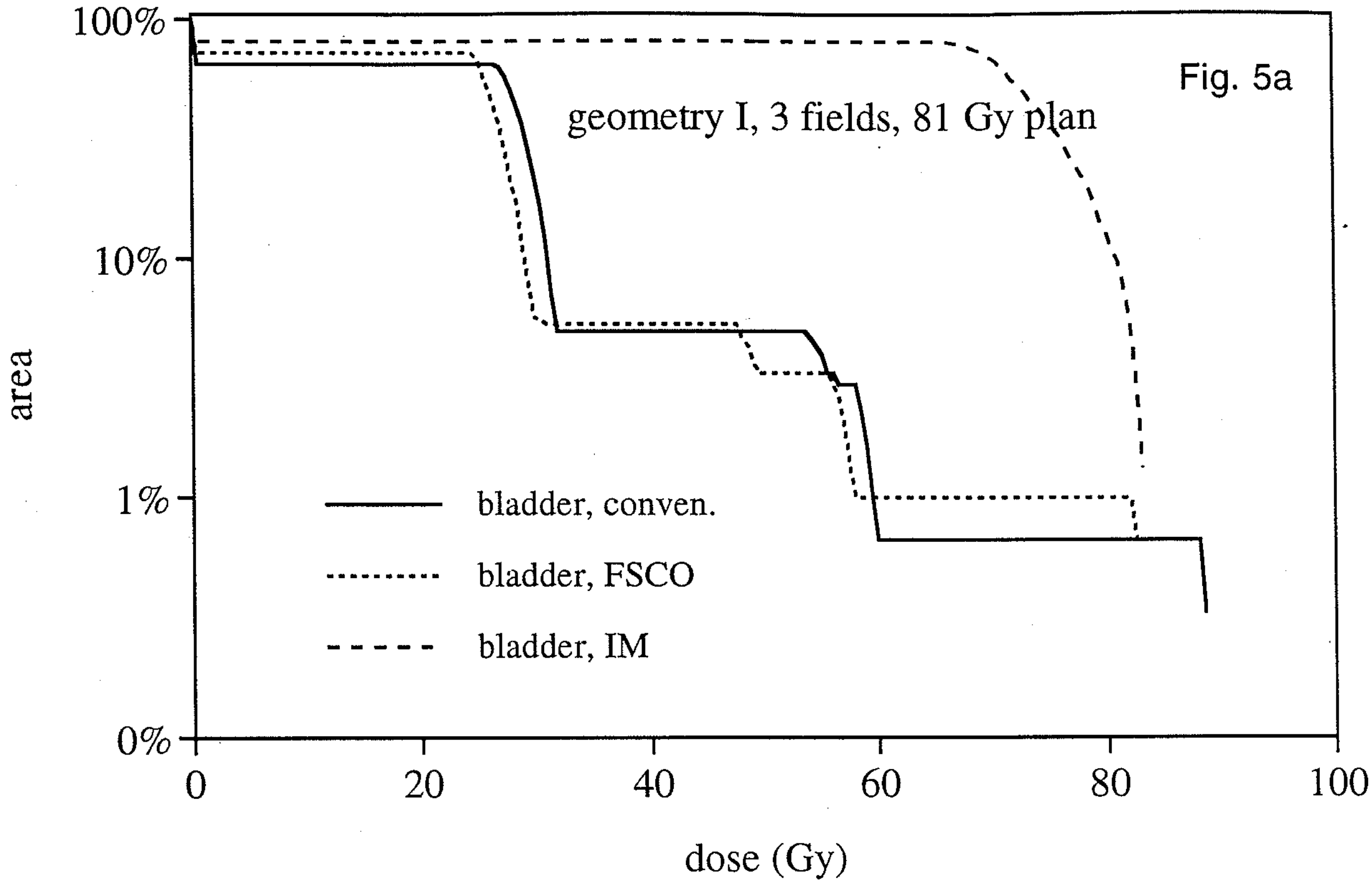

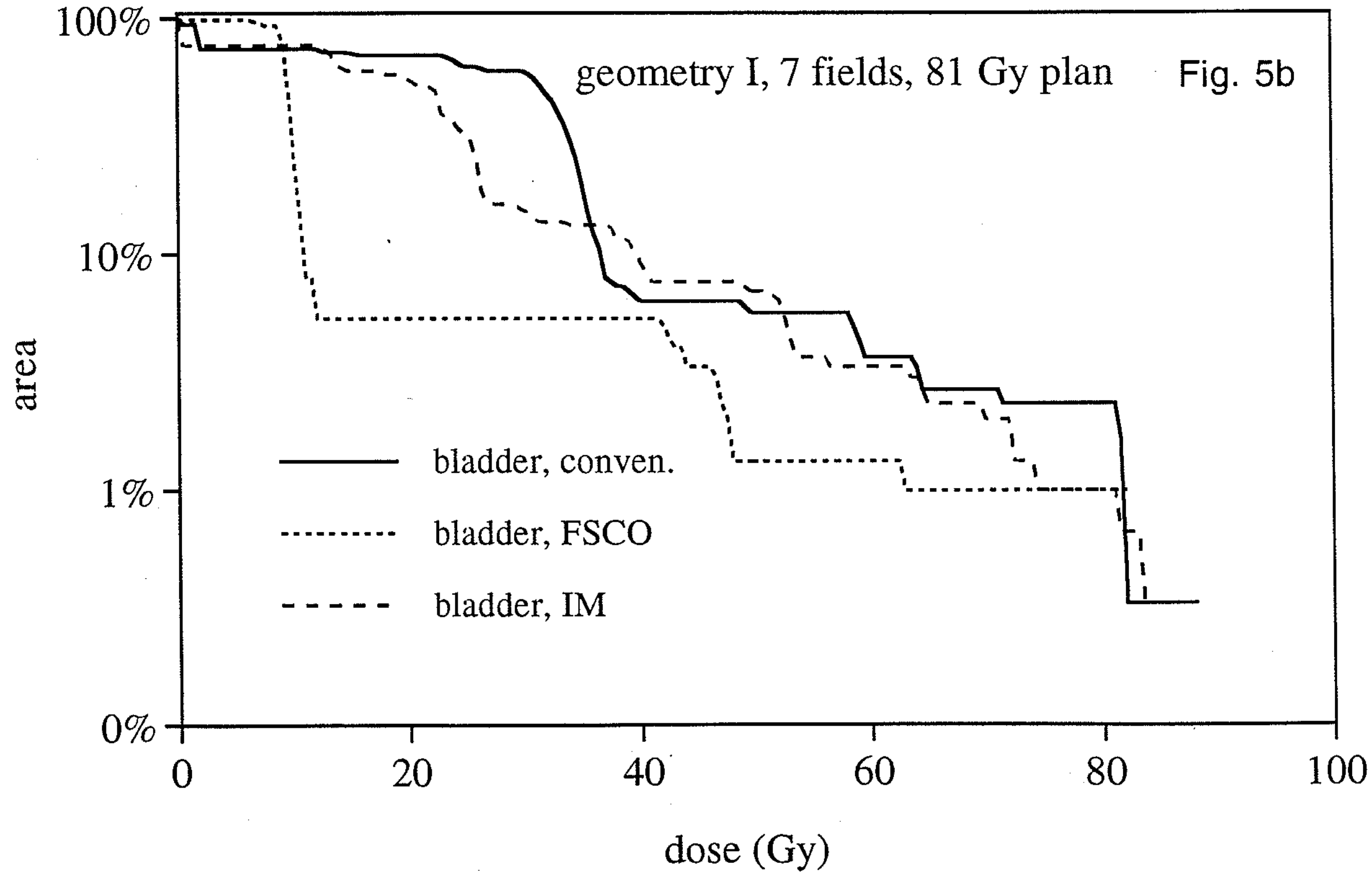

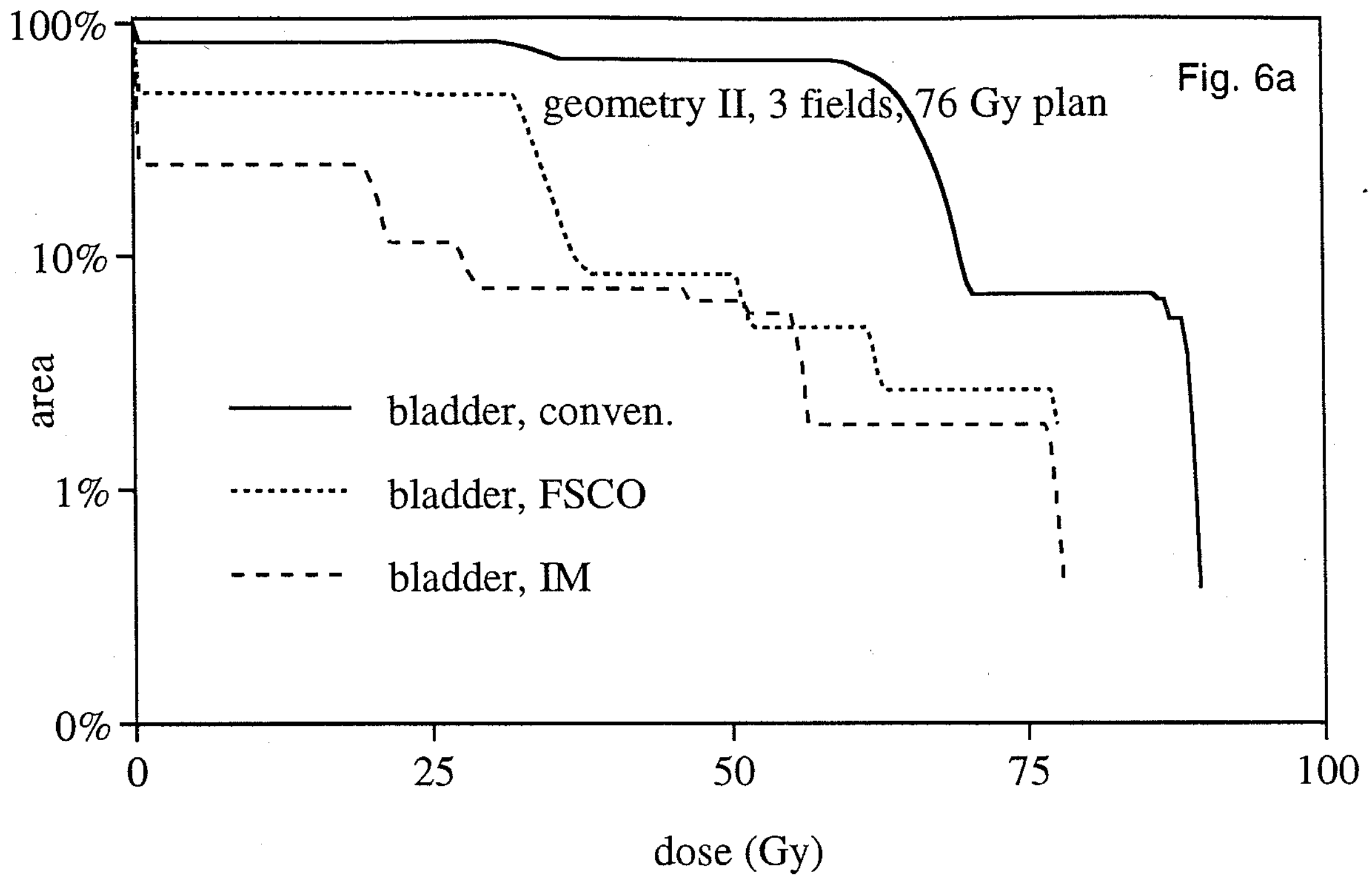

Fig. 6a

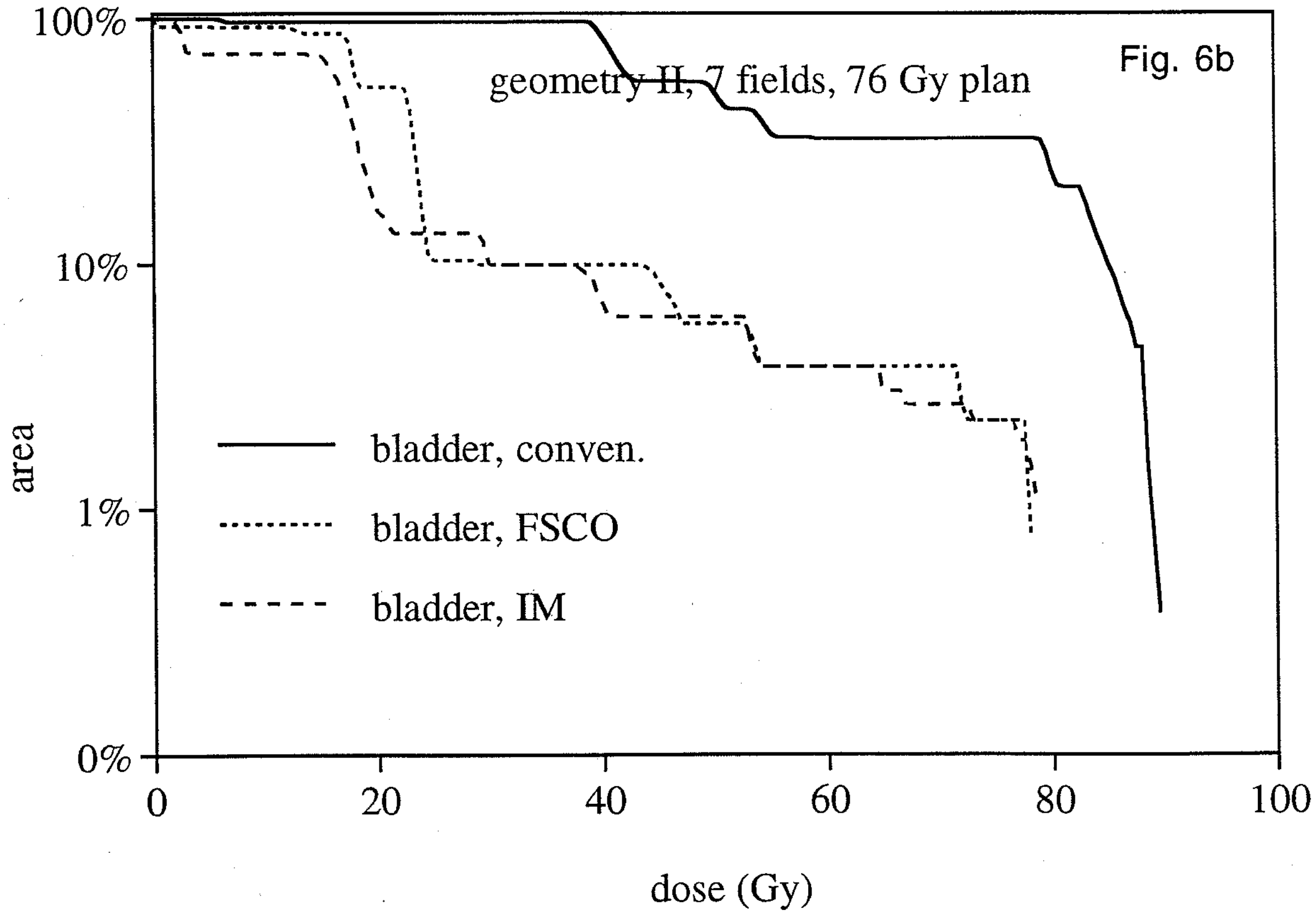

Fig. 6b

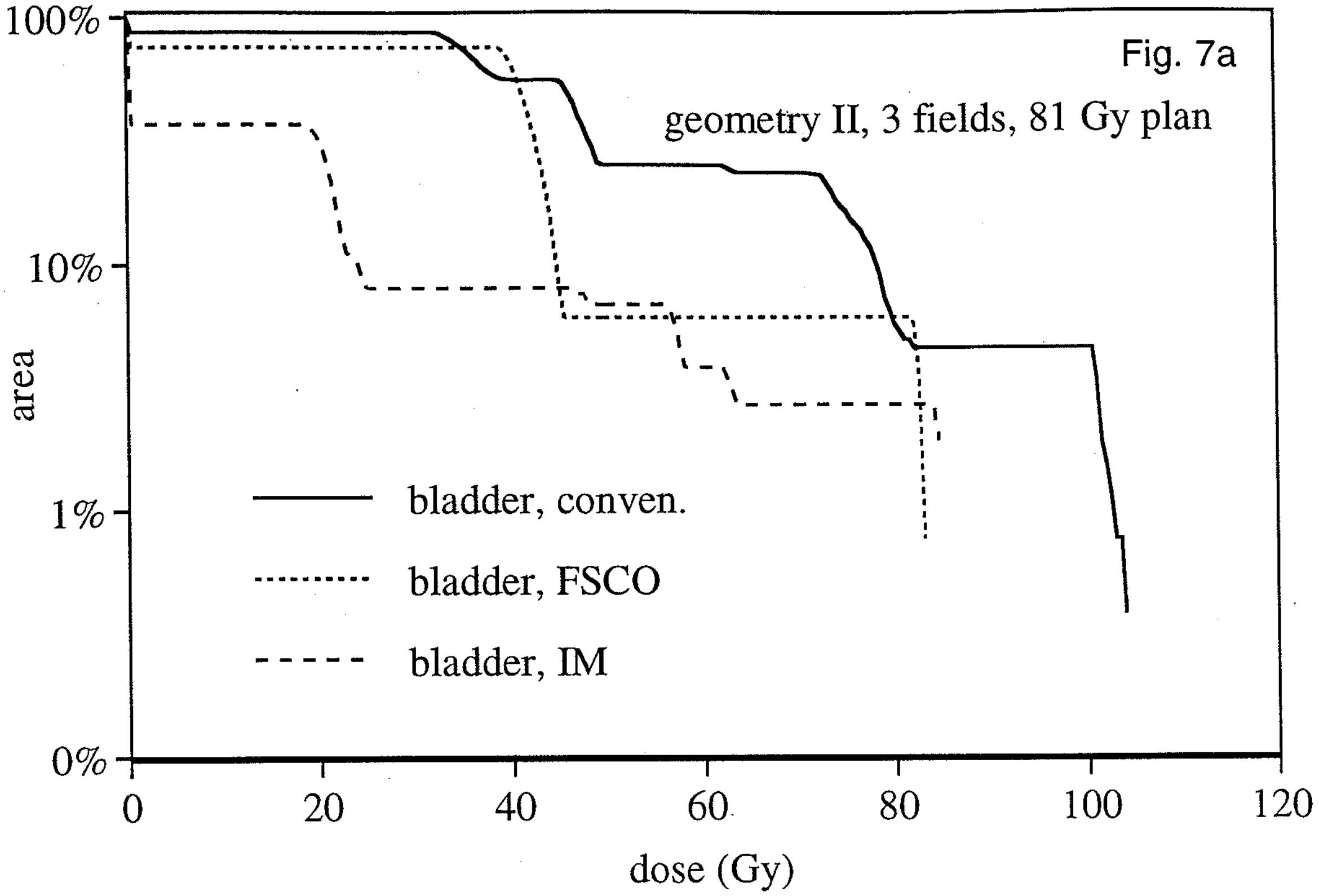

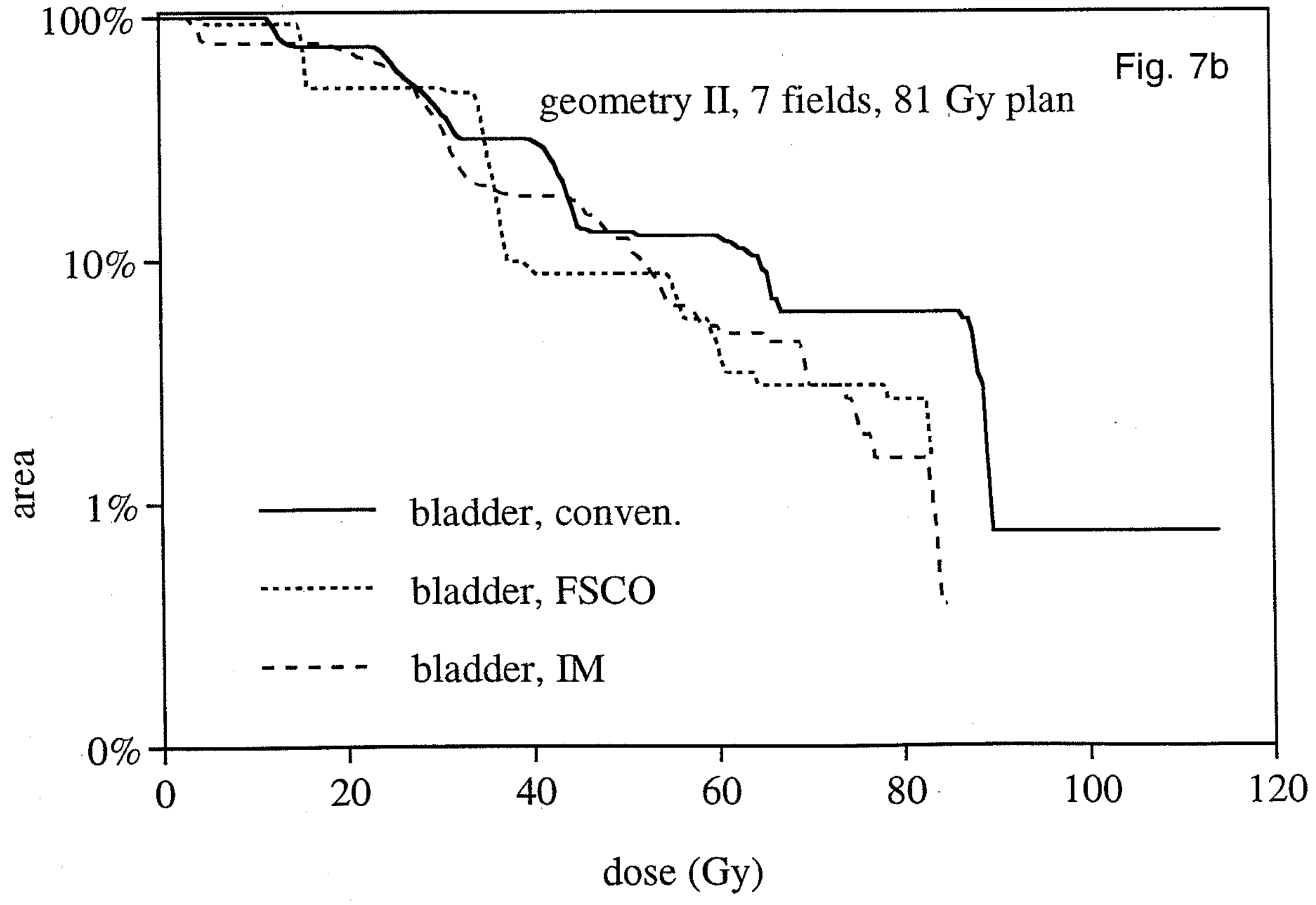

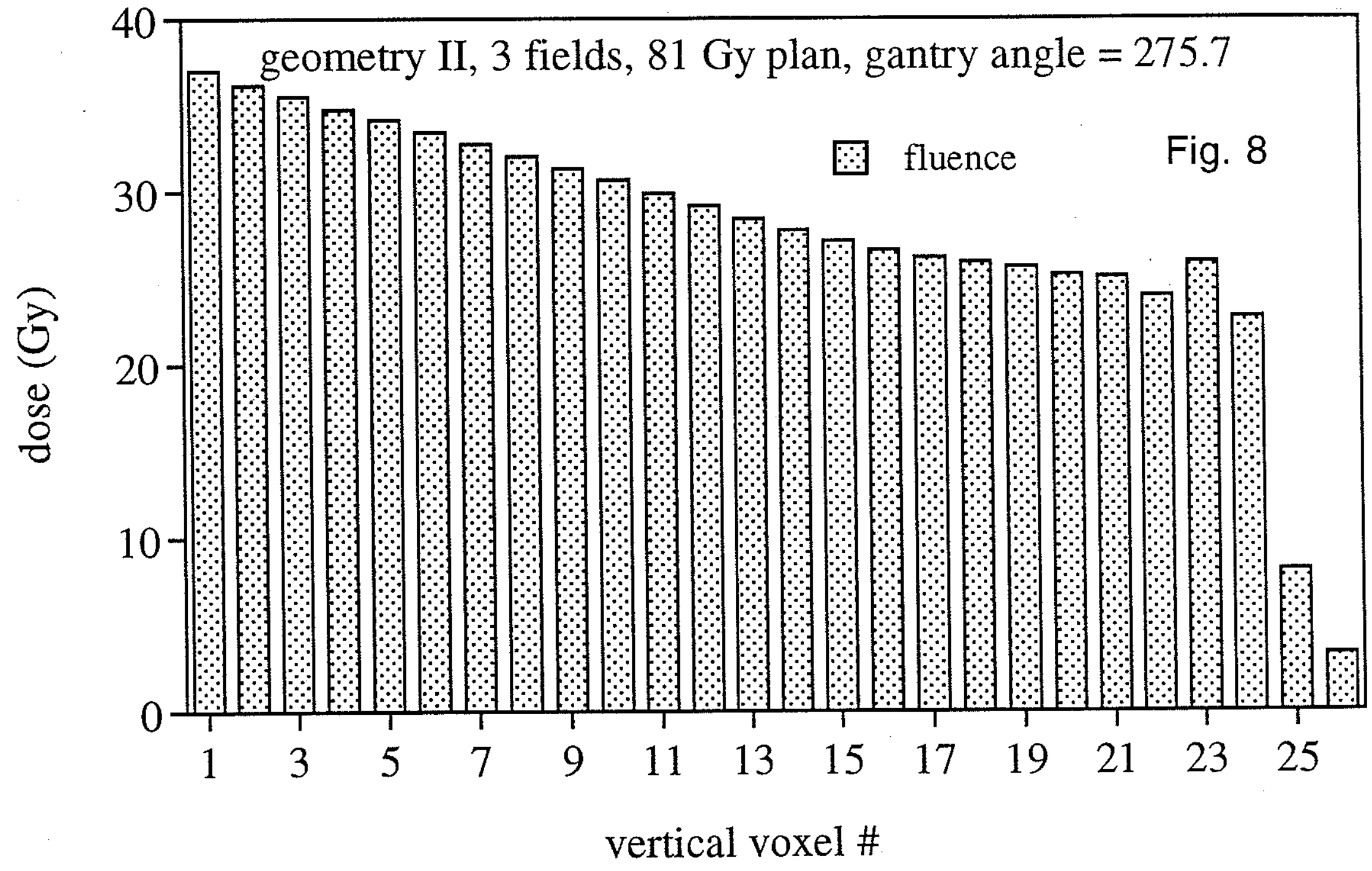
geometry II, 3 fields, 81 Gy plan, gantry angle = 275.7
fluence
Fig. 8
dose (Gy)
0
10
20
30
40
1
3
5
7
9
11
13
15
17
19
21
23
25
vertical voxel #

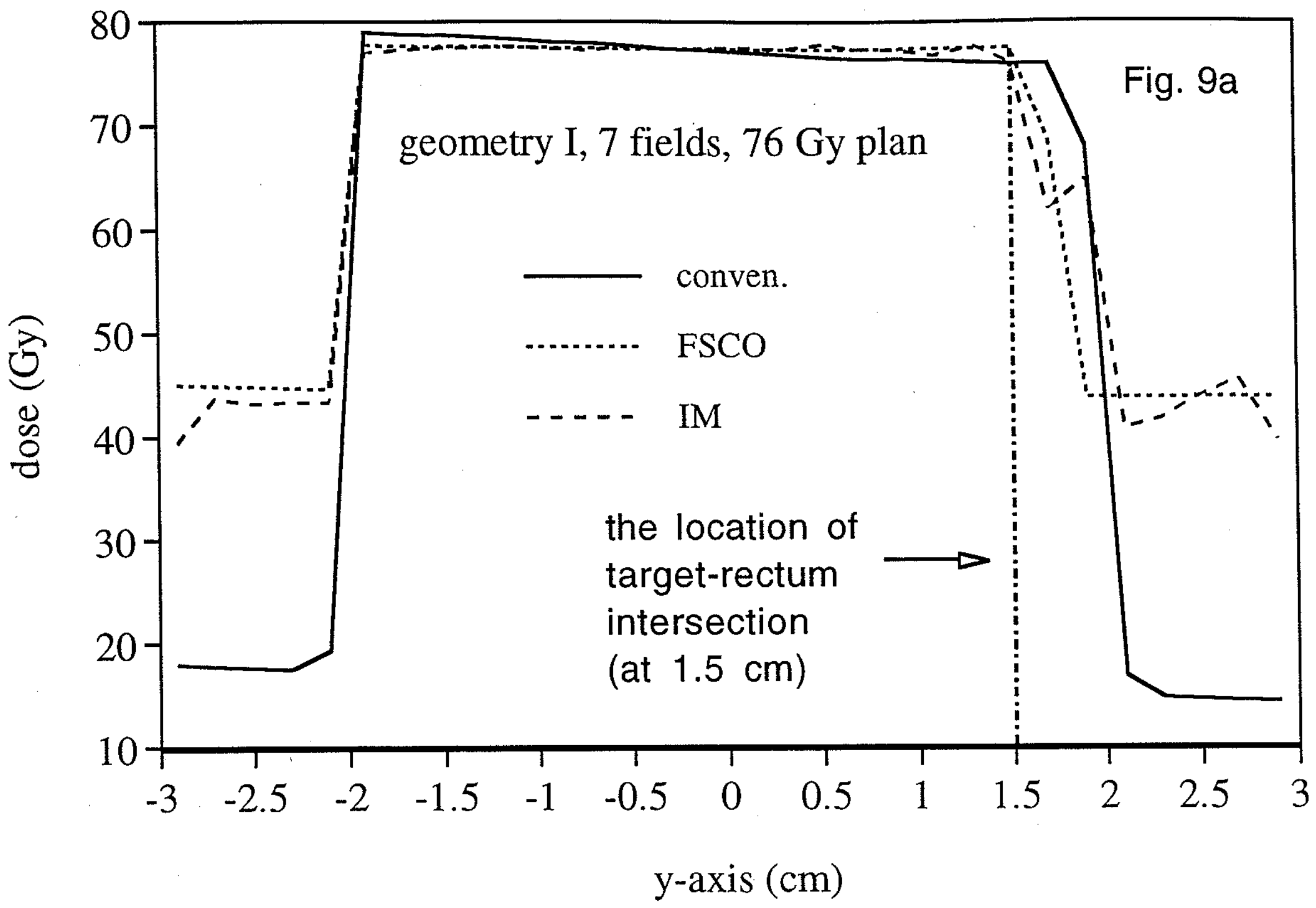

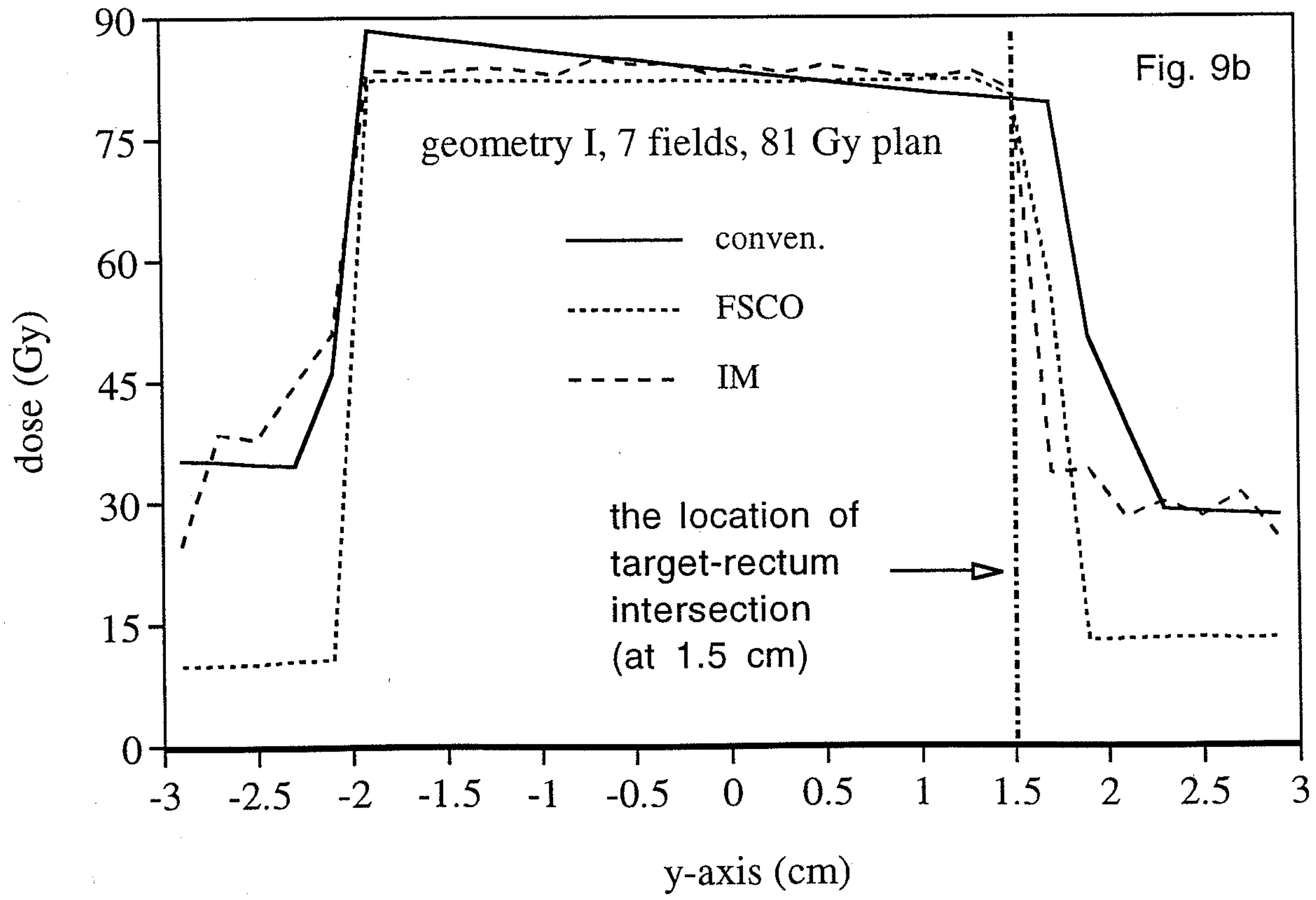

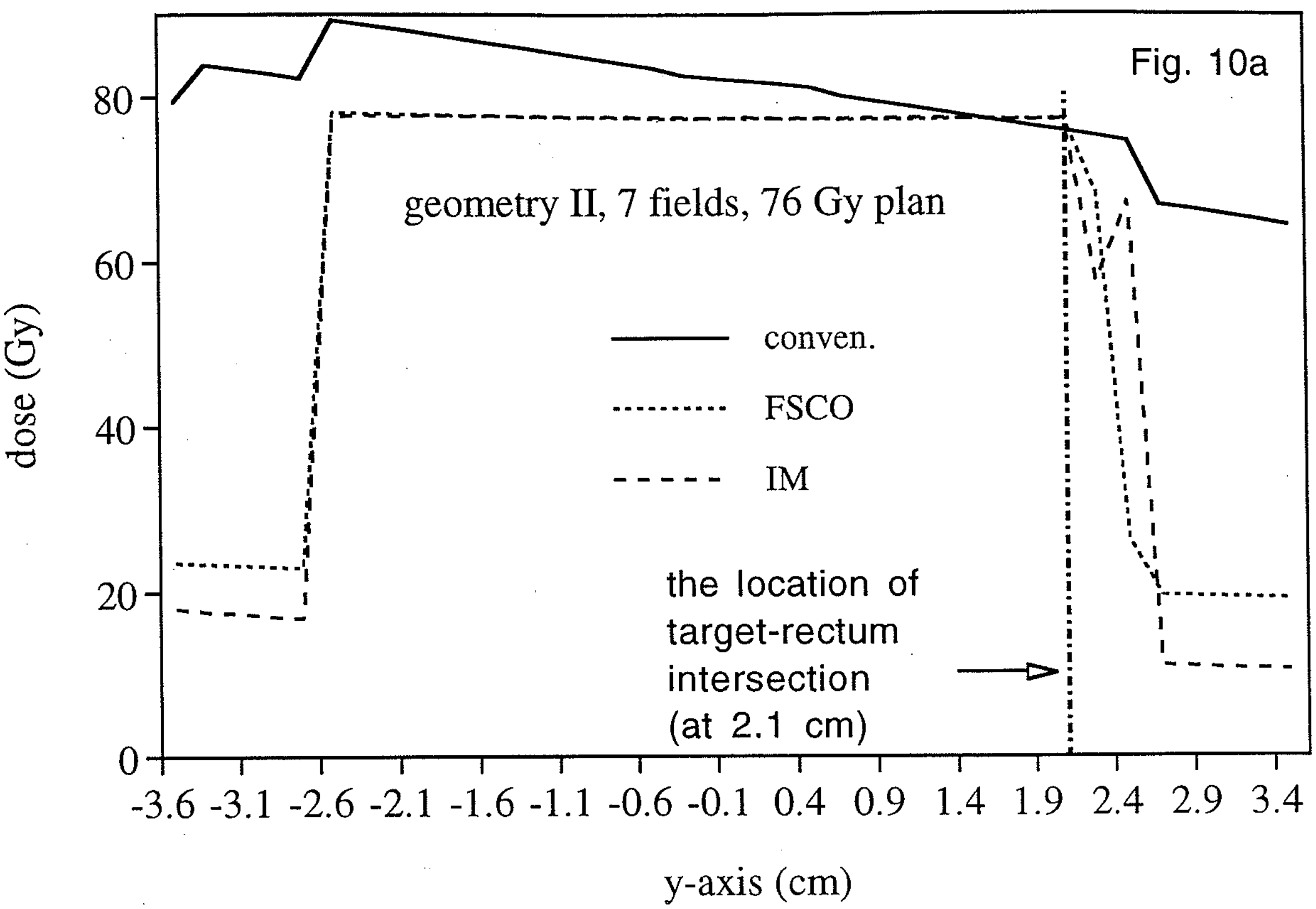

Fig. 10a
geometry II, 7 fields, 76 Gy plan
conven.
FSCO
IM
the location of
target-rectum
intersection
(at 2.1 cm)
dose (Gy)
0
20
40
60
80
-3.6 -3.1 -2.6 -2.1 -1.6 -1.1 -0.6 -0.1 0.4 0.9 1.4 1.9 2.4 2.9 3.4
y-axis (cm)

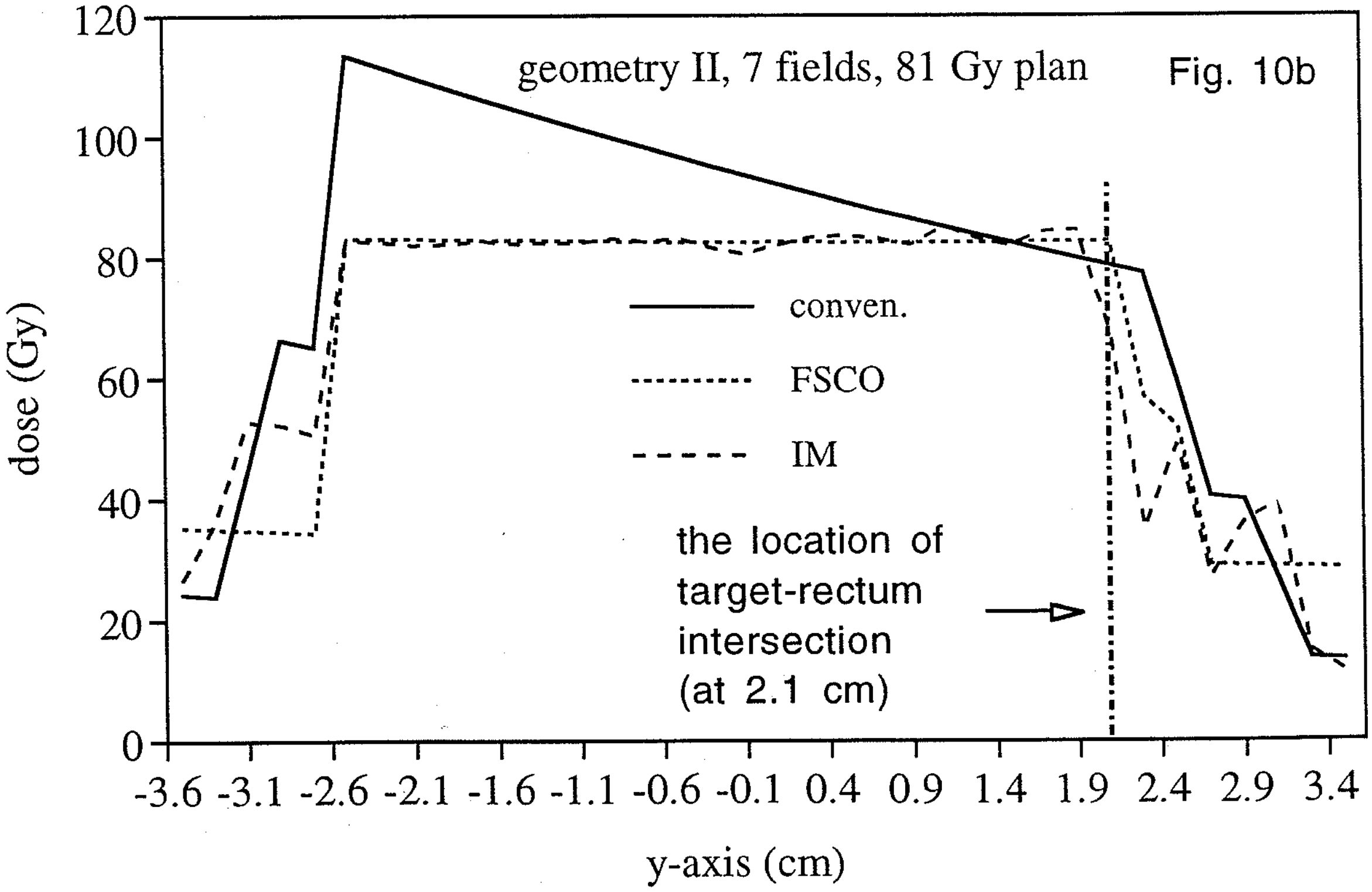

geometry II, 7 fields, 81 Gy plan
Fig. 10b
conven.
FSCO
IM
the location of
target-rectum
intersection
(at 2.1 cm)
dose (Gy)
0
20
40
60
80
100
120
-3.6 -3.1 -2.6 -2.1 -1.6 -1.1 -0.6 -0.1 0.4 0.9 1.4 1.9 2.4 2.9 3.4
y-axis (cm)

Fig. 11a: Beam Configuration

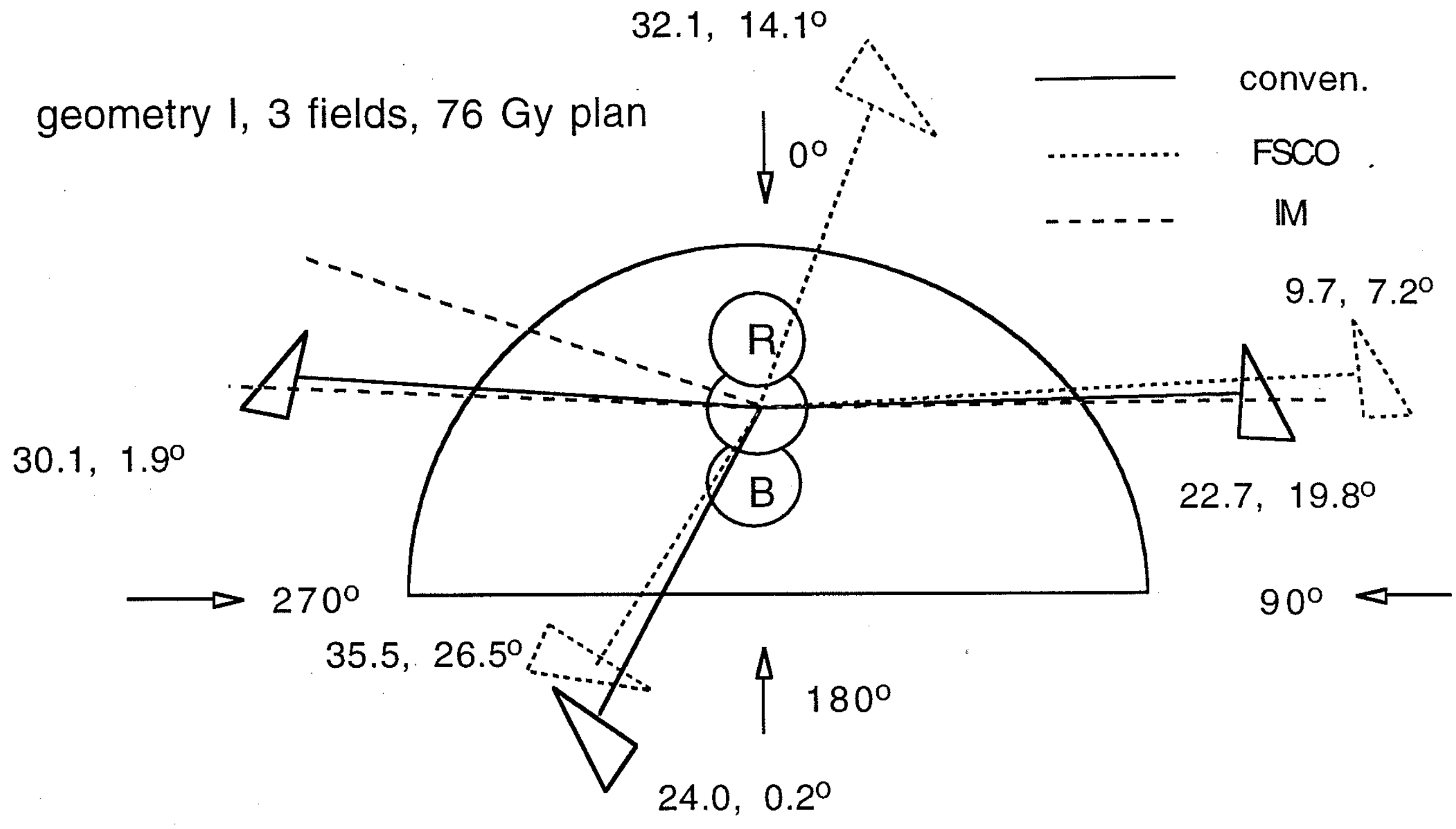

Fig. 11b: Beam Configuration

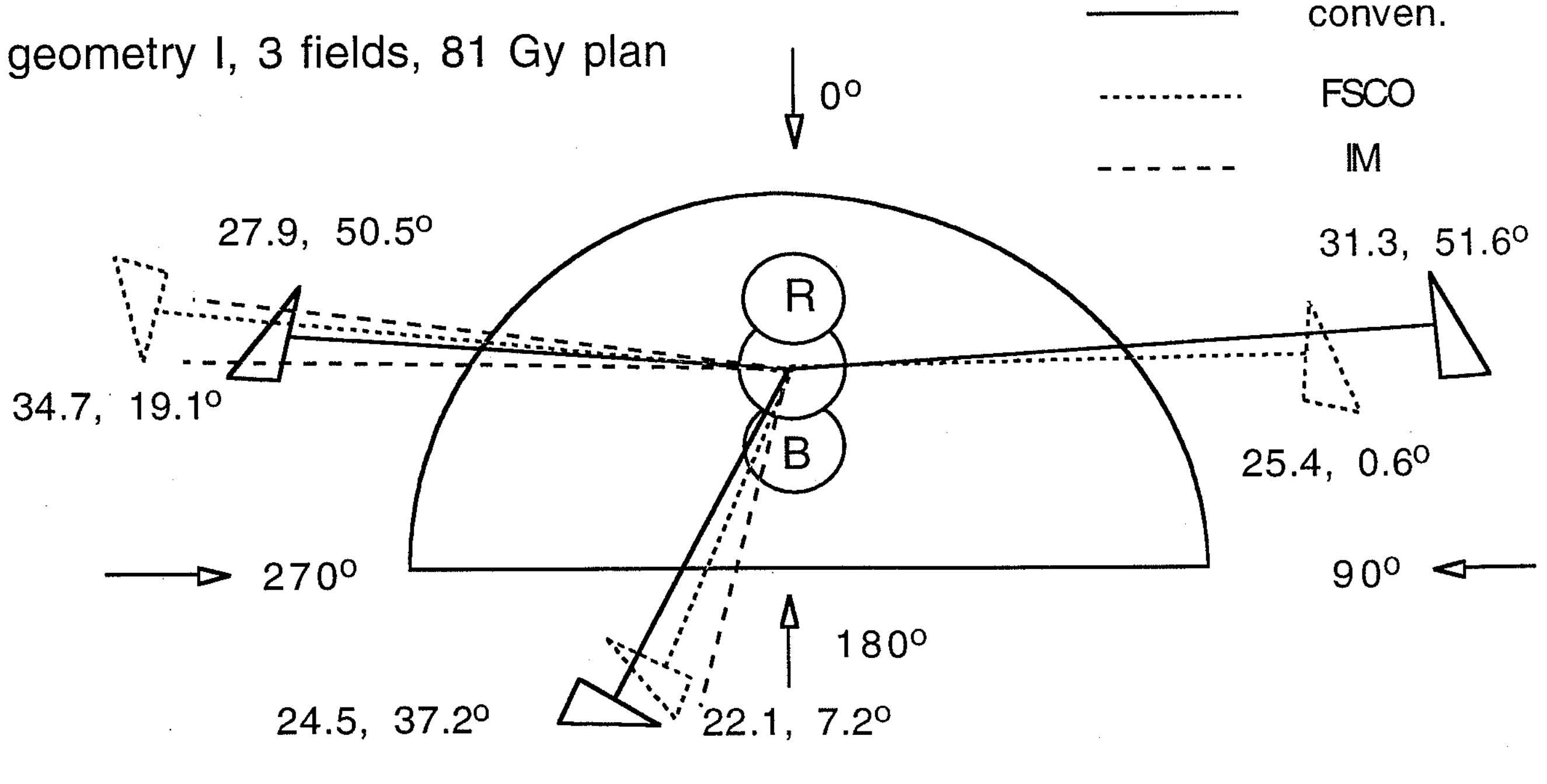

Fig. 12a: Beam Configuration

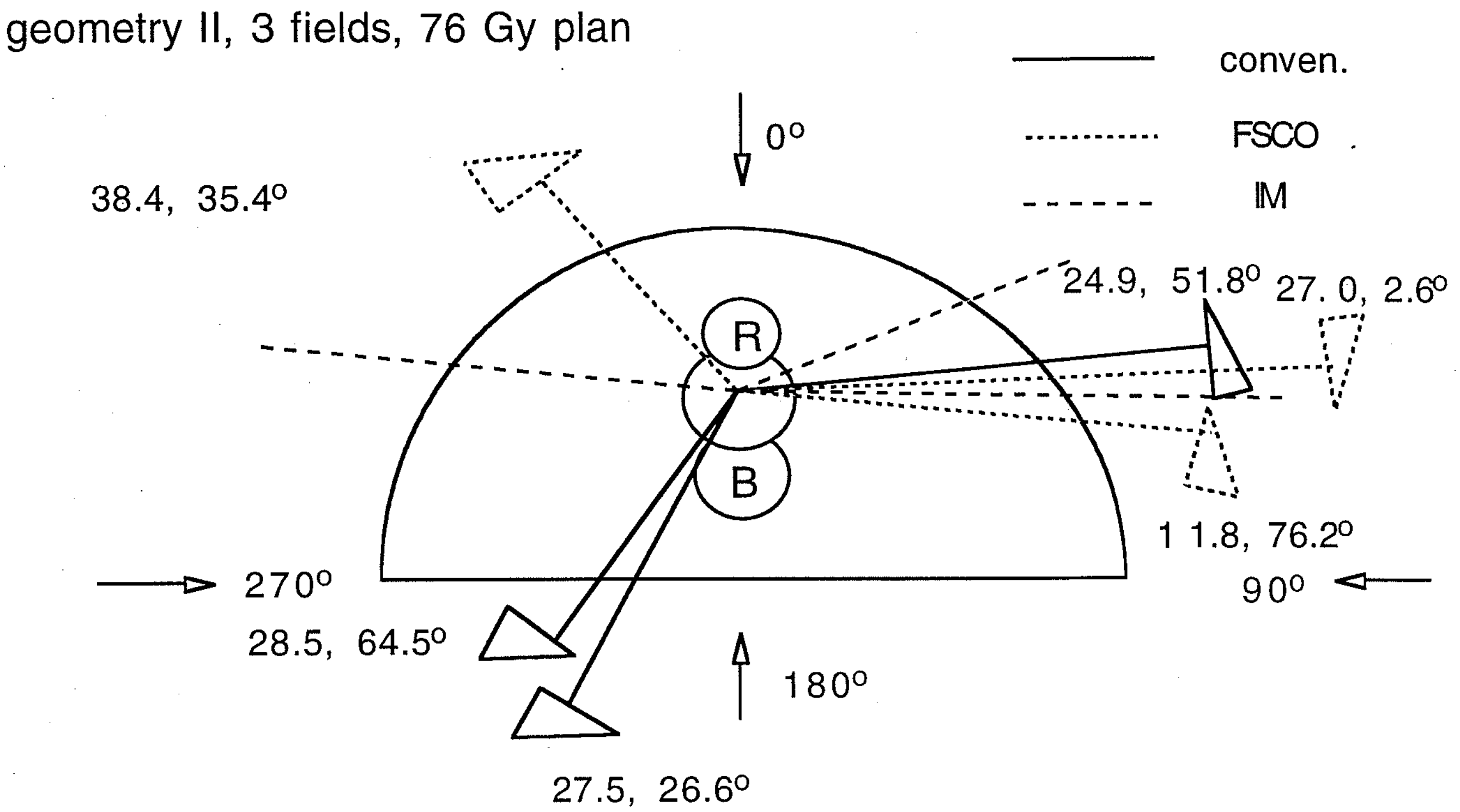

Fig. 12b: Beam Configuration

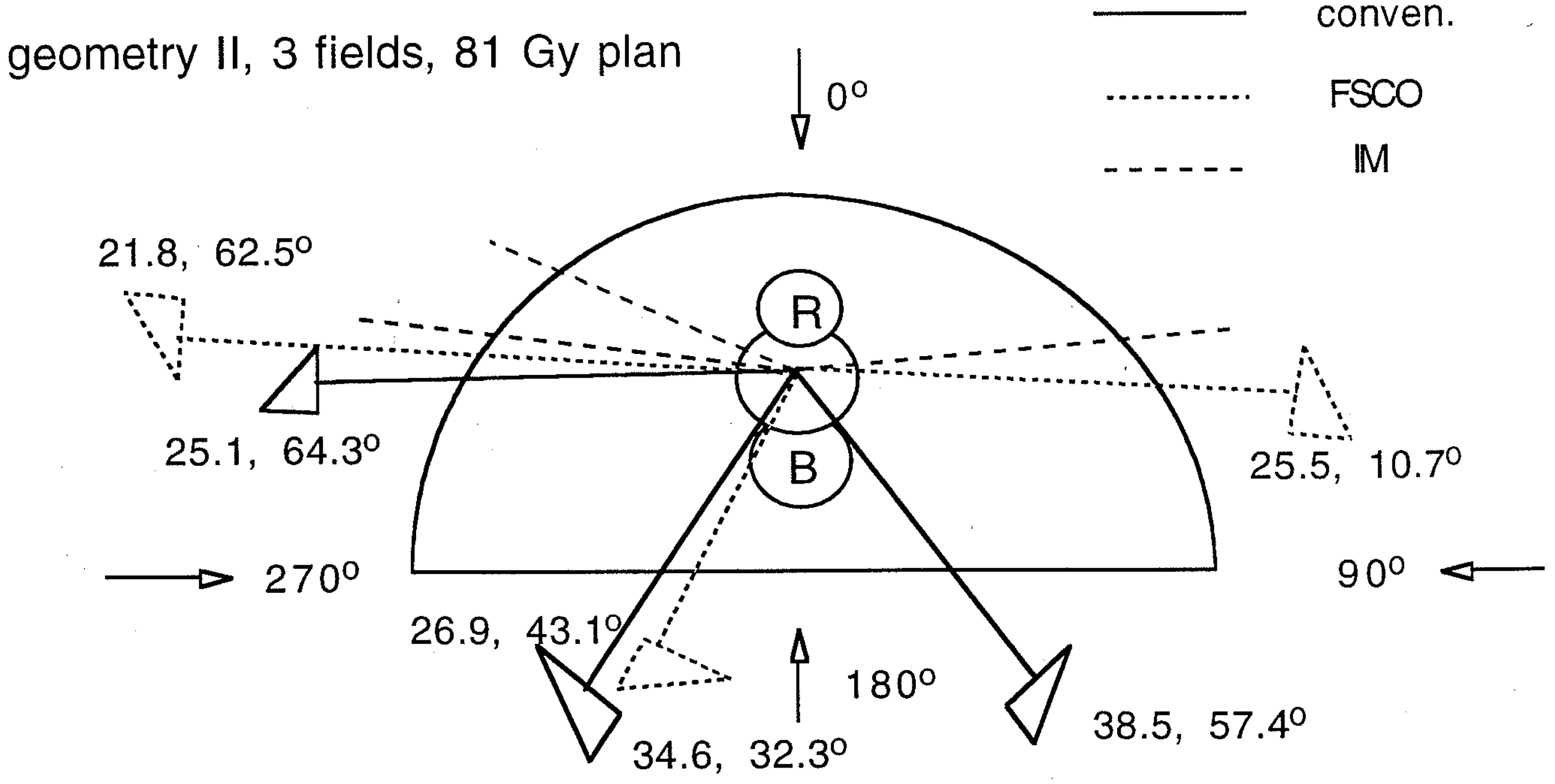